\documentclass[prb,aps,twocolumn,showpacs,amsmath,floatfix]{revtex4}
\usepackage{bbm}
\usepackage{amssymb}
\usepackage{amstext}
\usepackage{amscd}
\usepackage{amsthm}
\usepackage{amsopn}
\usepackage{amsmath}
\usepackage{amsbsy}
\usepackage{indentfirst}
\usepackage{graphicx}
\usepackage{verbatim}
\usepackage{bm}
\usepackage[bookmarks=false]{hyperref}

\begin{document}
\author{Sung-Po Chao, Vivek Aji}
\affiliation{Department of Physics and
Astronomy, University of California, Riverside, CA 92521}
 \title{Electric and thermoelectric transport in graphene and helical metal in finite magnetic fields}

\begin{abstract}
We study electrical and thermoelectric transport properties of the surface state of the topological insulator and graphene in the presence of randomly distributed impurities. For finite impurity strength, the dependence of the transport coefficients as a function of gate voltage, magnetic field and impurity potential, are obtained numerically. In the limit of zero impurities (clean limit), analytic results for the peak values of the magneto-oscillations in thermopower are derived. Analogous with the conventional two dimensional electron gas, the peak values are universal in the clean limit. Unlike graphene, in topological insulators the coupling of the electron spin to its momentum leads to a dependence of the transport coefficients on the gyromagnetic ratio ($g$).  We compare our results with data on graphene and identify unique signatures expected in topological insulators due to the magnetoelectric coupling.
\end{abstract}
\maketitle
\section{Introduction}
Helical Dirac fermions,  massless relativistic charged particles with spin locked to their linear momentum, are proposed to exist on the surface 
of three dimensional topological insulators (TIs)\cite{Kane,Roy,Fang} and later confirmed in experiments\cite{Hsieh,Xia} on samples like Bi$_2$Se$_3$ and Bi$_{0.9}$Sb$_{0.1}$. For compounds such as HgTe and Bi$_2$Se$_3$, surfaces with a single Dirac cone have been found\cite{Hsieh,Zhang}. While the low energy spectrum is very similar to that of graphene, the difference in the microscopic origin of the phenomena points to subtle differences. For example, the presence of a single Dirac cone in TIs, as opposed to two doubly degenerate ones in graphene, leads to significant reduction in phase space at low energies. Furthermore, the coupling of physical spin, rather than pseudospin (related to sublattice symmetry), in TIs suggests that the response in magnetic fields will be markedly different.

The key property of the surface states of TIs is the magnetoelectric effect. Unfortunately, the presence of bulk carriers has made it hard to observe it in transport measurements\cite{Ong,steinberg1,analytis1,check1,jchen, wei1}. One way to overcome this is to identify unique signatures, in transport, of the surface states. In particular the anomalous dependence of transport coefficients on the gyromagnetic ratio (g-factor) can be exploited to this end. The advantage of this approach is that the anomalous contribution can be enhanced by applying an in plane electric field\cite{Mondal}. In this article we study the thermopower and magnetotransport properties of these surface Dirac fermions in the presence of randomly distributed impurities. 

In the conventional two dimensional electron gas under a quantizing magnetic field, the diagonal thermopower, $S_{xx}$, of the clean sample at low temperatures shows a series of peaks near the Landau levels $N$. The peak values are $\ln2 k_B/e(N+1/2)$ independent of magnetic field strength\cite{Jonson}. One of our goals is to show how these universal features are modified in graphene and TIs.

For electrons exhibiting a  Dirac spectrum, such as  graphene,  this height of $S_{xx}$ is expected to vary as $\ln2 k_B/eN$, as required by the Berry phase effects leading to a $\frac{1}{2}$ integer shift in the Landau level index\cite{Zuev,Check}. We verify this result for $|N|\ge 1$ and discuss the singular behavior near $N=0$. We use self consistent Born approximation (SCBA) to consider the effect of short ranged, randomly distributed impurities on transport and magnetotransport\cite{Peres,Ando} of the Dirac fermions and compute diagonal and off diagonal thermopower numerically. Similar to the results in conventional 2D metals, the height of $S_{xx}$ at low temperature does not have universal value after the inclusion of impurities\cite{Jonson}.

We also show that for intermediate impurity strength, not considered in previous studies which have focused on the unitarity limit\cite{Peres, Ando, Ugarte}, the particle hole symmetry of the original Dirac spectrum is broken under SCBA while for weak and strong impurity strength the particle hole symmetry is restored. Another source of scattering in graphene is charged Coulomb impurities and its effect on conductivity and thermopower, both within perturbation\cite{Hwang, Dassarma} and SCBA\cite{Yan, Ugarte}, have been extensively studied. Comparison to available data\cite{Zuev,Check,Wei} suggests that a single scattering mechanism cannot account for all observed features\cite{Ugarte}. Our formulation for TIs, with the g-factor set to zero, is similar to the case of graphene. Thus, in addition to exploring the difference in the transport properties induced by the spin orbit coupling, we can also compare our prediction with the available data. Qualitative agreement with thermopower data is obtained for all but the zeroth Landau level. The failure is related to the underestimation of the longitudinal conductivity.

The important new feature in topological insulators, as compared to graphene, is the spin-orbit coupling. In addition to orbital quantization, an external magnetic field couples to the momentum via its Zeeman coupling with the spin. Such an interaction reveals itself in novel signatures in transport unique to topological insulators. For example, the universal amplitudes in thermopower depend on the gyromagnetic ratio ($g$), but no splitting of peaks occurs as the surface states are derived from a single Dirac cone. While the dimensional coupling constant $\alpha=g^2\mu_B^2/ v_F^2e$ is quite small for $HgTe$ (of order $10^{-4}/T$) to yield a measurable signature, we derive the expected behavior in a model system where $\alpha$ is enhanced by an order of magnitude. The enhancement is achievable in principle by applying an inplane electric field\cite{Mondal}. 

The paper is organized as follows: in Sec.~\ref{s2} we describe the general Hamiltonian and T-matrix formulation used to perturbatively compute the effects of impurities. In Sec.~\ref{s3} we present our results, within linear response, for conductivity and the diagonal thermopower $S_{xx}$ in the absence of magnetic fields. In Sec.~\ref{s4} we consider the linear response conductivity and the thermopower $S_{xx}$ and $S_{xy}$ in the presence of the magnetic field. We compare our numerical results with those obtained in the graphene experiments and other theoretical results. Conclusion are summarized in Sec.~\ref{s5}.  

\section{Hamiltonian and $T$ matrix formulation}\label{s2}  
The surface state of the electrons on a topological insulator under electric and magnetic fields is\cite{Mondal}
\begin{eqnarray}\label{ho}
H_0=\int d\vec{r} \psi^{\dagger}(\vec{r})[v_F\vec{\sigma}\cdot\vec{\pi}-\mu I-g\mu_B\vec{\sigma}\cdot\vec{B}-eE x] \psi(\vec{r})
\end{eqnarray}
where $v_F$ is the Fermi velocity ($\hbar$ is restored in computing current current correlation), $\vec{\sigma}=\sigma_{x}\hat{x}+\sigma_y\hat{y}$ denotes the Pauli matrices in spin space, $I$ is the $2\times 2$ identity matrix, $\psi(\vec{r})=(\psi_{\uparrow},\psi{\downarrow})^T$ is the annihilation operator for Dirac spinor ($T$ denotes the transpose of a row vector)and  $\vec{\pi}=-i\vec{\nabla}+e\vec{A}$ is the canonical momenta. 
Electric field is assumed to be pointed in x-direction and magnetic field is perpendicular to the plane (z-direction). The coupling to local impurities of strength $V$ is modeled as
\begin{eqnarray}\label{hi}
H_{imp}=\int d\vec{r} \sum_{\vec{r}_i,s}V\delta(\vec{r}-\vec{r}_i)\psi^{\dagger}_s(\vec{r})\psi_s(\vec{r})
\end{eqnarray}
where $s$ denotes spin degree of freedom of electron operator $\psi(\vec{r})$ and sum over $\vec{r}_i$ indicates summing all impurity positions.  
The full Hamiltonian of the system is given by $H=H_0+H_{imp}$.
In the dilute impurity limit we compute transport coefficients to linear order in impurity density $n_i=N_i/N$. The finite temperature Green's functions of the electrons are 
\[
 G_{s,s'}(\vec{k},\vec{p},\tau )=\left(\begin{array}{cc}
G_{\uparrow,\uparrow}(\vec{k},\vec{p},\tau ) & G_{\uparrow,\downarrow}(\vec{k},\vec{p},\tau ) \\
G_{\downarrow,\uparrow}(\vec{k},\vec{p},\tau ) & G_{\downarrow,\downarrow}(\vec{k},\vec{p},\tau )  \end{array}\right)\] 

\noindent where $\vec{k}=(k_x,k_y)$ is the two dimensional momentum of the surface band, $s$ and $s'$ are the spin indices, and 
\[G_{s,s'}(\vec{k},\vec{p},\tau)=-\langle {\bf T}\psi_{\vec{k},s}(\tau)\psi_{\vec{p},s'}(0)\rangle\]
with ${\bf T}$ being the time ordering operator. To linear order in $n_i$, the Greens function satisfy
\begin{eqnarray}
\mathbb{G}(\vec{k},\omega_n)=\mathbb{G}^0(\vec{k},\omega_n)+\mathbb{G}^0(\vec{k},\omega_n)\mathbb{T}(\omega_n)\mathbb{G}(\vec{k},\omega_n)
\end{eqnarray}
where $\omega_n=2\pi(n+1/2)/\beta$ is the fermionic Matsubara frequency and $$\mathbb{G}(\vec{k},\omega_n)=\frac{1}{\beta}\int d\tau e^{i\omega_n\tau} G_{s,s'}(\vec{k},\vec{k},\tau).$$

The $T$ matrix $\mathbb{T}(\omega_n)$ is 
\begin{eqnarray}\label{tmatrix}
 \mathbb{T}(\omega_n)=Vn_i[1-\frac{V}{N}\sum_{\vec{k}}\mathbb{G}^0(\vec{k},\omega_n)]^{-1}
\end{eqnarray}
In the following two sections we discuss the nature of transport without and with the magnetic field. We use the above T-Matrix formulation to compute the Greens function. Transport coefficients are obtained within linear response using the appropriate current-current correlation function.

\section{Zero magnetic field}\label{s3}
\begin{figure}[t]
\includegraphics[width=.5\columnwidth, clip]{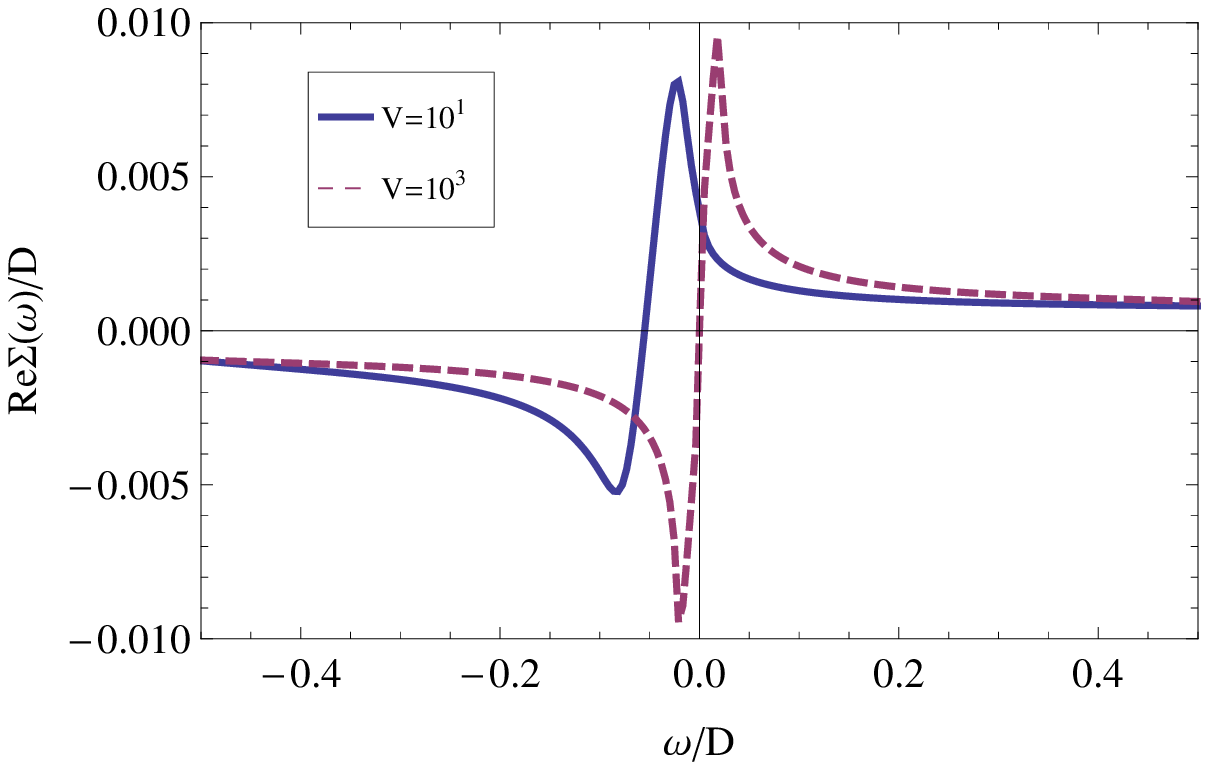}\hfill
\includegraphics[width=.5\columnwidth, clip]{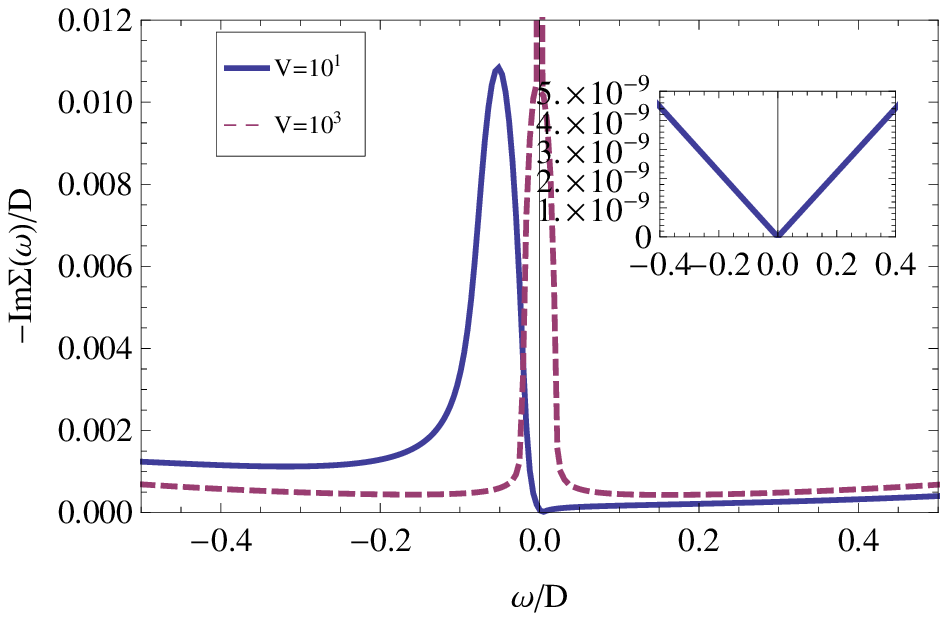}
\caption{Real part (left) and imaginary part (right) of the self energy plot. $D=3eV$, $n_{imp}=10^{-3}$, and $V=10eV$ for blue lines and $V=10^3eV$ for dashed purple lines. The inset of right figure shows imaginary part of self energy for weak impurity potential $V=10^{-2}eV$.}
	  \label{figf1}
\end{figure}

In zero magnetic field and $E\simeq 0$ the Hamiltonian for the helical metal without impurities reduces to
\[
H_0=\int d\vec{r} \psi^{\dagger}(\vec{r})v_F\left(\begin{array}{cc}
-\mu & k_x-ik_y\\
k_x+ik_y & -\mu \end{array}\right)\psi(\vec{r})\]
The eigenfunction of the Schordinger equation $$H_0 \int d\vec{r} F_{\vec{k},\gamma}(\vec{r})c^{\dagger}_{\vec{k},\gamma}|0\rangle=(\gamma v_F k-\mu)\int d\vec{r}F_{\vec{k},\gamma}(\vec{r})c^{\dagger}_{\vec{k},\gamma}|0\rangle$$ in the energy eigenstate is given by $\int d\vec{r}F_{\vec{k},\gamma}(\vec{r})c^{\dagger}_{\vec{k},\gamma}|0\rangle$. Here $|0\rangle$ is the vacuum, $c^{\dagger}_{\vec{k},\gamma}$ denotes spinless fermion operator, $|\vec{k}|=k$ , and the spin part is described by the two component spinor function $F_{\gamma,\vec{k}}(\vec{r})$ is
\begin{equation}\label{sp1}
 F_{\gamma,\vec{k}}(\vec{r})=\frac{1}{\sqrt{2 A}}e^{i\vec{k}\cdot\vec{r}}\left( \begin{array}{c}
\gamma \\ e^{i\phi(\vec{k})} \end{array}\right)
\end{equation}
with $\gamma=+/- 1$ denoting conduction/valence bands, $\phi(\vec{k})=\tan^{-1}(k_y/k_x)$, and $A$ is the area of the system. Following the full self consistent Born approximation\cite{Peres} we have the self energy of electron given by
\begin{eqnarray}\label{sp2}
\Sigma(\omega+i0^+)=\frac{V n_i}{1-\frac{V}{N}\sum_{\vec{k},\gamma}[\omega+\mu-\gamma v_F k-\Sigma(\omega)]^{-1}}
\end{eqnarray}
The derivation of Eq.(\ref{sp2}) is in the Appendix.\ref{A0}.
This self energy is solved numerically as shown in Fig.\ref{figf1}. The main feature of this self energy is that for moderate impurity strength ($V=10eV$) the particle hole symmetry is not preserved, while for strong ($V=10^3eV$) or weak ($V=10^{-2}eV$) impurity interaction the particle hole symmetry is restored. The scale is set by the bandwidth which in our calculations is $6 eV$.

The physical quantity we are interested in is the transport properties of surface states. In the Kubo formulation the linear response conductivity is related to current current correlation. Written in Matsubara formulation the particle current current correlator is\cite{Jonson}
\begin{eqnarray}\label{kubo}
L_{ij}^{11}(i\omega)=\frac{-iT}{\hbar(i\omega)N\bar{V}}\int_0^{\hbar\beta}d\tau e^{i\omega\tau}\langle J_i(\tau)J_j(0)\rangle
\end{eqnarray}
Here $N\bar{V}=A$ is the area of the system, with $N$ being the number of unit cells and $\bar{V}$ denoting unit cell area. $L_{ij}^{11}$ is related to the linear response charge conductivity $\sigma_{ij}$ via 
\begin{eqnarray}\label{cond}
L_{ij}^{11}=(T/e^2)\sigma_{ij}
\end{eqnarray}

The particle current operator in $k$ direction in the Heisenberg representation is
\begin{eqnarray}\label{particlecurrent}
J_k=\frac{\partial r_k}{\partial t}=i[H,r_k]=-v_F\sigma_k
\end{eqnarray}
Note that the electron current is related to the spin of the electron and hence sensitive to the Zeeman coupling. The corresponding particle current operator in the energy eigenbasis, Eq.(\ref{sp1}), is 
\begin{eqnarray}\nonumber
&&J_x=-\frac{v_F}{2}\sum_{\gamma,\bar{\gamma},\vec{k}}(\gamma e^{i\phi(\vec{k})}+\bar{\gamma}e^{-i\phi(\vec{k})})c_{
\vec{k},\gamma}^{\dagger}c_{\vec{k},\bar{\gamma}}\\\label{fc}
&&J_y=i\frac{v_F}{2}\sum_{\gamma,\bar{\gamma},\vec{k}}(\gamma e^{i\phi(\vec{k})}-\bar{\gamma}e^{-i\phi(\vec{k})})c_{
\vec{k},\gamma}^{\dagger}c_{\vec{k},\bar{\gamma}}
\end{eqnarray}
Within the limit of dilute random impurities we use the SCBA which is valid as long as $k_F l\gg 1$ (with $l$ as electron's mean free path)\cite{Bruus}.
The Kubo formula for the real part of particle current current correlation is then given as
\begin{eqnarray}\nonumber
L_{xx}^{11}(\omega,T,\mu)=\frac{Tv_F^2}{\hbar\omega N\bar{V}}\int\frac{d\epsilon}{2\pi}[n_F(\epsilon)-n_F(\epsilon+\omega)]\\\times\sum_{\vec{k}}\Im G(\vec{k},\epsilon+i0^+)\Im G(\vec{k},\epsilon+\omega+i0^+)
\end{eqnarray}
with $n_F(\epsilon)=1/(e^{\beta\epsilon}+1)$  (Fermi Dirac distribution function) and $L^{11}_{xy}=0$. Quantum effects such as weak localizations are not considered in this article as the major contributions to the conductivity comes from non crossing diagrams. The momentum integral $k$ is cut off by linear spectrum energy boundary $D$ with $k_c v_F=D$. This upper cutoff is also related\cite{Peres} to unit cell size $\bar{V}$ by $\pi k_c^2=(2\pi)^2/\bar{V}$.

\begin{figure}[t]
\includegraphics[width=.5\columnwidth, clip]{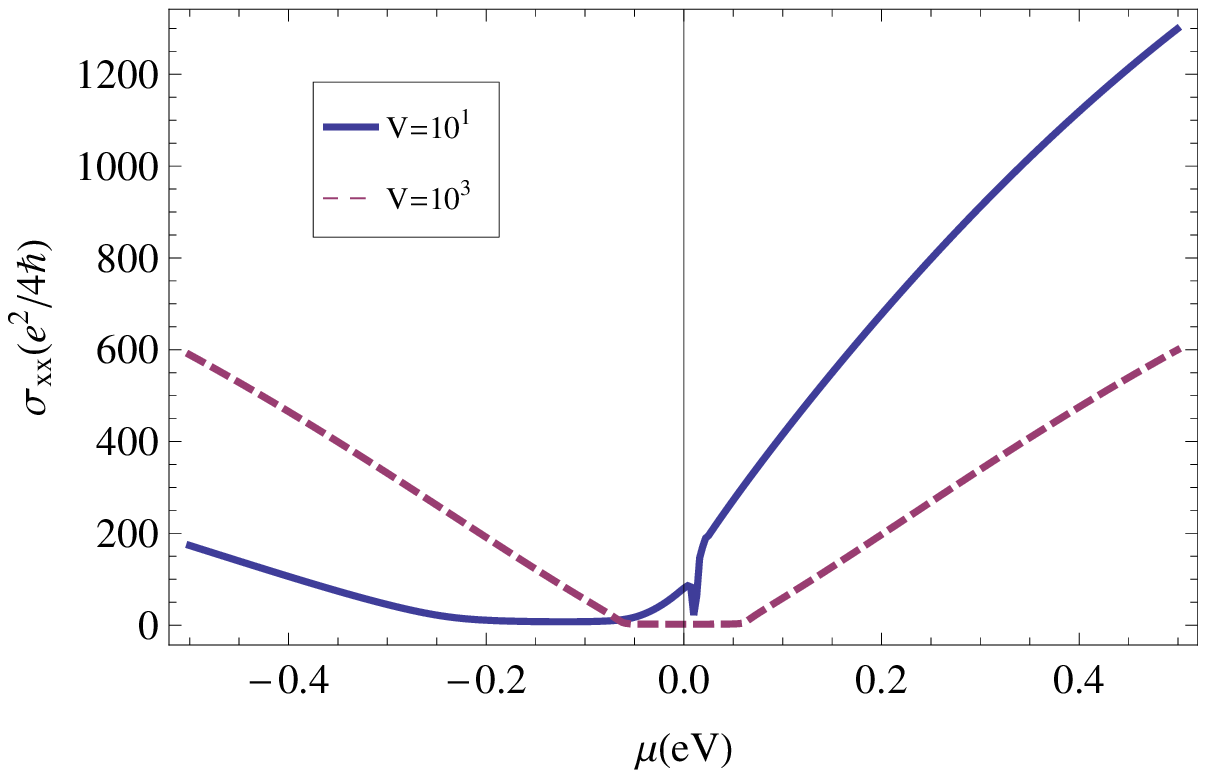}\hfill
\includegraphics[width=.5\columnwidth, clip]{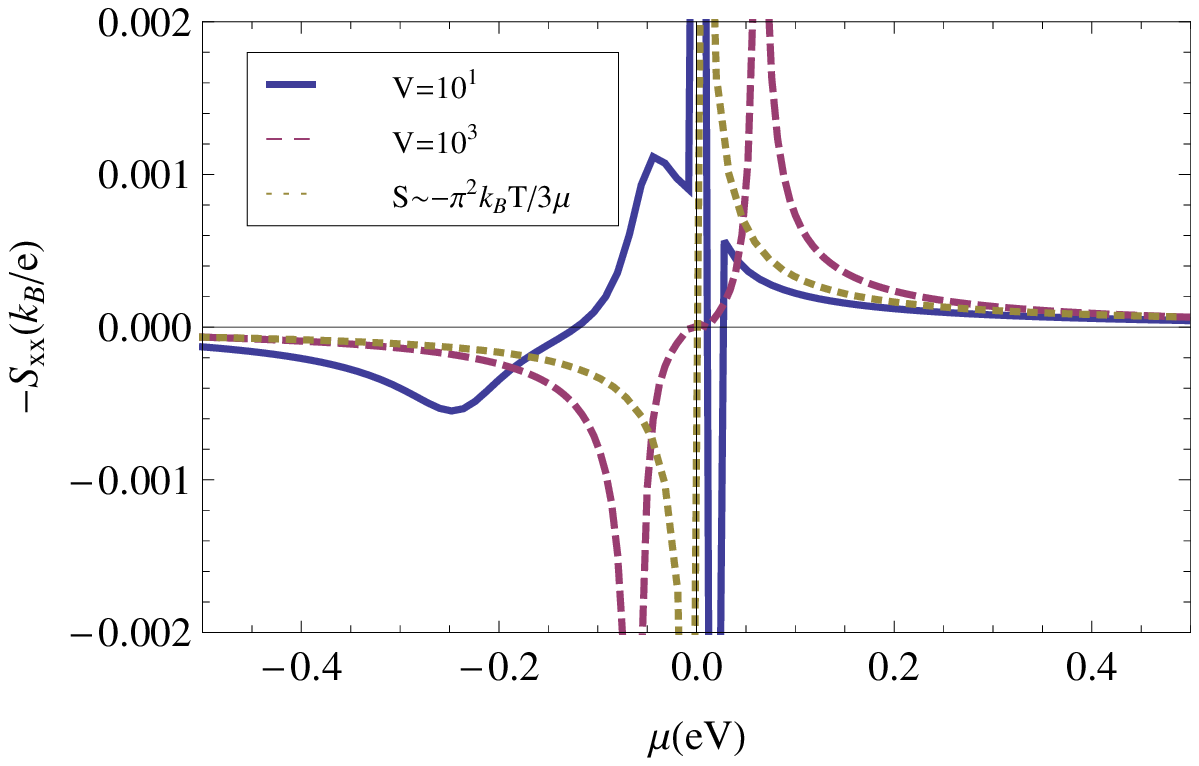}
\caption{DC conductivity $\sigma_{xx}(e^2/4\hbar)$ (left) and thermopower $S_{xx}(k_B/e)$ (right) at $k_BT=10^{-5}eV$ vs chemical potential $\mu(eV)$ for $V=10$ (blue lines) and $V=10^3eV$ (dashed purple lines). The conductivity at $\mu=0$ for dashed purple line is around $e^2/2\pi h$. The dotted brown line on the right figure is the first term in Eq.(\ref{cls}) which is the dominant thermopower in large $\mu$.}
	  \label{figf2}
\end{figure}

The zero frequency thermal response function $L_{ij}^{12}$ is related to $L_{ij}^{11}$ by\cite{Jonson}
\begin{eqnarray}\label{l12}
&&\lim_{\omega\rightarrow 0}L_{ij}^{12}(\omega,T,\mu)\\\nonumber&&=\lim_{\omega\rightarrow 0}\int_{-\infty}^{\infty}d\epsilon \epsilon\frac{-\partial n_F(\epsilon)}{\partial\epsilon}L_{ij}^{11}(\omega,T=0,\epsilon+\mu),
\end{eqnarray}
and the thermopower $S_{ij}$ is 
\begin{eqnarray}\label{tp}
S_{ij}=\sum_{m}(-1/eT)(L^{11})^{-1}_{im}L_{mj}^{12}
\end{eqnarray}
In zero magnetic field only diagonal component of $S_{ij}$ is nonzero. For a given self energy $\Sigma(\epsilon)=\Re\Sigma(\epsilon)+i\Im\Sigma(\epsilon)$ the DC conductivity is 
\begin{eqnarray}
&&\sigma_{xx}(\mu,T)=\frac{-e^2}{2\pi h}\int d\epsilon\frac{\partial n_F(\epsilon-\mu)}{\partial\epsilon}\\\nonumber&&\times\int_0^D d(v_Fk)(v_Fk)(\frac{\Im\Sigma(\epsilon)}{(\epsilon-\Re\Sigma(\epsilon)-v_Fk)^2+(\Im\Sigma(\epsilon))^2}\\\nonumber
&&+\frac{\Im\Sigma(\epsilon)}{(\epsilon-\Re\Sigma(\epsilon)+v_Fk)^2+(\Im\Sigma(\epsilon))^2})^2\\\nonumber
&&=\frac{-e^2}{2\pi h}\int d\epsilon\frac{\partial n_F(\epsilon-\mu)}{\partial\epsilon}\kappa(\epsilon,V,n_{imp})
\end{eqnarray}
The explicit form of $\kappa(\epsilon,V,n_{imp})$ is given in the Appendix \ref{A1}. Notice that for $\mu\simeq 0$ the low temperature conductivity is proportional to $\kappa(\mu\simeq0,V,n_{imp})\simeq 1$. Thus the low temperature conductivity at Dirac node has an universal value $e^2/2\pi h$. Close to zero temperature Eq.(\ref{cond}) to Eq.(\ref{tp}) lead to the generalized Mott formula for the thermopower\cite{Jonson}
\begin{eqnarray}\label{mott}
S_{ij}=-\frac{\pi^2k_B^2T}{3e}\sum_m[\sigma^{-1}]_{im}[\partial\sigma/\partial\mu]_{mj}
\end{eqnarray}
For a clean surface state we take $V\simeq 0$ and $n_{imp}\simeq 0$ and thus away from the half filling we have $D\gg\mu-\Re\Sigma(\mu)\gg\Im\Sigma(\mu)\simeq 0$.
Near zero temperature in the clean limit the thermopower takes the form
\begin{eqnarray}\nonumber
S_{xx}&\simeq&-\frac{\pi^2k_B}{3e}\left(\frac{1-\partial_{\mu}\Re\Sigma(\mu)}{(\mu-\Re\Sigma(\mu))/k_BT}-\frac{\partial_{\mu}\Im\Sigma(\mu)}{\Im\Sigma(\mu)/k_BT}\right)\\\label{cls}&\simeq&-\frac{\pi^2k_B}{3e}\left(\frac{1}{\mu/k_BT}-\frac{\partial_{\mu}\Im\Sigma(\mu)}{\Im\Sigma(\mu)/k_BT}\right)
\end{eqnarray}
Since $\frac{\partial_{\mu}\Im\Sigma(\mu)}{\Im\Sigma(\mu)/k_BT}$ in general is nonzero, the thermopower in the clean limit is susceptible to impurity interaction and does not show universal behavior for a given chemical potential, different from what occurs in the case of finite magnetic field which we 
show in the next section. The DC conductivity and diagonal thermopower for general impurity strength are computed numerically and shown in Fig.(\ref{figf2}).
Similar to the results of self energy in Fig.(\ref{figf1}) the DC conductivity shows asymmetry in $\mu$ for moderate impurity strength. The dip near $\mu=0$ for 
$V=10eV$ case is the remnant signature of the clean sample, which shows up as a singularity in thermopower of the form $S_{xx}\propto\frac{-k_BT}{\mu}$ for $\mu\simeq 0$.
\section{Finite magnetic field}\label{s4}
We assume the magnetic field $\vec{B}$ is perpendicular to the applied electric field $\vec{E}=E\hat{x}$ and choose the gauge $\vec{A}=(0,B x \sin\theta,-Bx\cos\theta)$. For $\vec{E}=0$ the eigenvalue of $H_0$ is given by the Landau level spectrum
\begin{eqnarray}\nonumber
E(\gamma,n)&=&-\mu+\gamma v_F l_B^{-1}\sqrt{2n+\alpha B \sin\theta} \\\nonumber
&=&-\mu+\gamma \epsilon_n  \mbox{     for $n\subset \bf{N}$}\\
E(\gamma,0)&=&-\mu-g\mu_B B\sin\theta =-\mu+\epsilon_0
\end{eqnarray} 
where $\alpha=g^2\mu_B^2/ v_F^2e$, $\gamma=\pm$ is the band index, $l_B=\sqrt{1/eB\sin\theta}$ is the magnetic length, and  $n$ denotes the index of Landau level. For Dirac electrons on the surface of HgTe $\alpha\simeq 10^{-4}/T$. Thus for usual magnetic field strength we can in general ignore the Zeeman contribution in the spectrum \cite{Mondal}. Nevertheless we keep this Zeeman contribution in our calculation and the dominant feature that results from the Zeeman term is the lack of particle hole degeneracy at $n=0$ level. The loss of symmetry leads to anomalous features in off diagonal conductivity $\sigma_{xy}$ as well as thermopower. Within linear response we take $\vec{E}\sim 0$ in the Hamiltonian and the field operator $\psi(\vec{r})$ are expressed in linear combinations of energy eigenstates.
\begin{eqnarray}\nonumber
&&\psi(\vec{r})
 =\sum_k\frac{e^{i(k+g\mu_B B\cos\theta) y}}{\sqrt{L}}\left(\begin{array}{c} 0\\ \phi_{k,0}(x)  \end{array}\right)c_{k,0}
\\&&+\sum_{n,k,\gamma}\frac{e^{i(k+g\mu_B B\cos\theta)y}}{\sqrt{(1+f_{n,\gamma}^2)L}}
\left(\begin{array}{c} \phi_{k,n-1}(x) \\ \gamma f_{\gamma,n}\phi_{k,n}(x)\end{array}\right)c_{k,n,\gamma}
\end{eqnarray}
with $f_{\gamma,n}=(\gamma\epsilon_n+g\mu_B B \sin\theta)/(\gamma v_F\sqrt{2n}/l_B)$,
\begin{eqnarray}\nonumber
\phi_{k,n}(x)=\frac{i^n \exp\left(\frac{-1}{2}\left(\frac{x+l_B^2 k}{l_B}\right)^2 \right)}{\sqrt{2^n n!\sqrt{\pi}l_B}}H_n\left(\frac{x+l_B^2 k}{l_B} \right)
\end{eqnarray}
and $H_n(x)$ is the Hermite polynomial\cite{Peres,Ando}. The sum over level index $n$ is cut off by the linear spectrum boundary $E(\gamma,n_{max})=-\mu+\gamma D$. For $D\sim3.5eV$, $v_F\sim5\times10^5$ in the typical surface state of topological insulator\cite{Ong} the upper cutoff obtained this way is $n_{max}\simeq 10^6$. In our numerical computation we take $n_{max}=2000$ for practical computation\cite{Ando} and checked that the results do not change much by comparing with $n_{max}=3000$. In this representation $H_0$ is diagonal and the Green's function in energy eigenstate is 
\begin{eqnarray}
G_0(k,n,\gamma;i\omega)=\frac{1}{i\omega-E(\gamma,n)}
\end{eqnarray}
The impurity Hamiltonian, Eq.(\ref{hi}), in this basis is
\begin{widetext}
\begin{eqnarray}\nonumber
&&H_{imp}=\frac{V}{L}\sum_{i=1}^{N_{imp}}\sum_{p,k}e^{i(k-p)y_i}\Big[\phi_{p,0}^{\ast}(x_i)\phi_{k,0}(x_i)c^{\dagger}_{p,0}c_{k,0}+\sum_{n,\gamma}\frac{\gamma f_{\gamma,n}}{\sqrt{1+f_{\gamma,n}^2}}\phi_{p,0}^{\ast}(x_i)\phi_{k,n}(x_i)c^{\dagger}_{p,0}c_{k,n,\gamma}\\\nonumber&&+\sum_{m,\bar{\gamma}}\frac{\bar{\gamma} f_{\bar{\gamma},m}}{\sqrt{1+f_{\bar{\gamma},m}}}\phi^{\ast}_{p,m}(x_i)\phi_{k,0}(x_i)c_{p,m,\bar{\gamma}}^{\dagger}c_{k,0}
+\sum_{n,m,\gamma,\bar{\gamma}}\frac{1}{\sqrt{(1+f_{\gamma,n}^2)(1+f_{\bar{\gamma},m}^2)}}[\phi_{p,m-1}^{\ast}(x_i)\phi_{k,n-1}(x_i)\\&&+\gamma\bar{\gamma}f_{\bar{\gamma},m}f_{\gamma,n}\phi_{p,m}^{\ast}(x_i)\phi_{k,n}(x_i)]c^{\dagger}_{p,m,\bar{\gamma}}
c_{k,n,\gamma}\Big]
\end{eqnarray}
\end{widetext}
To include the effect of dilute random impurities we use the self consistent Born approximation and the averaging procedure over impurities positions as in Ref.~\onlinecite{Peres}. The position averaged Green's function is 
\begin{eqnarray*}
&&\langle G(k,n,\gamma;i\omega;\vec{r}_1,\vec{r}_2,\vec{r}_3,\ldots,\vec{r}_{N_{imp}})\rangle\\&&=\Big[\prod_{i=1}^{N_{imp}}\int \frac{d\vec{r}_i}{L^{2N_{imp}}}\Big] G(k,n,\gamma;i\omega;\vec{r}_1,\vec{r}_2,\vec{r}_3,\ldots,\vec{r}_{N_{imp}})\\&&\equiv G(k,n,\gamma;i\omega)
\end{eqnarray*}
In the averaging procedure we first integrate $y_i$ and then $x_i$ and use the following identity
\begin{eqnarray*}
\int dx \phi_{k,n}^{\ast}(x)\phi_{k,m}(x)=\delta_{n,m}
\end{eqnarray*}
and then sum over momentum index in the propagator by using
\begin{eqnarray*}
\int dk \phi_{k,n}^{\ast}(x)\phi_{k,m}(x)=\delta_{n,m}/l_B^2
\end{eqnarray*}
The self energy within self consistent Born approximation is 
\begin{eqnarray}\nonumber
&&\Sigma=n_i\Sigma_0({\bf 1}-\Sigma_0\mathbb{G})^{-1}\\\label{SCBA}
&&\mathbb{G}=(\mathbb{G}_0^{-1}-\Sigma)^{-1}
\end{eqnarray} 
In Eq.(\ref{SCBA}) the self energy, identity matrix ${\bf 1}$, and Green's functions are all of dimension $(2n_m+1)\times (2n_m+1)$
with $n_m=n_{max}$ as the maximal Landau level cutoff by the linear spectrum. Denote $$V_{\gamma,-\gamma,n}\equiv\frac{V(1-f_{\gamma,n}f_{-\gamma,n})}{\sqrt{(1+f_{\gamma,n}^2)(1+f_{-\gamma,n}^2)}}$$ 
and ${\bf\tilde{1}}_{n\times n}$ as $n$ by $n$ matrix with matrix element $1$ along $\{n_j,-n_j\}$,
the bare self energy term written in the energy eigenvalue basis is then given by
\[\Sigma_0=\left(\begin{array}{ccc}
V {\bf 1}_{n_m\times n_m} & 0 & V_{-,+,n} {\bf \tilde{1}}_{n_m\times n_m}\\
0 & V  &0 \\
 V_{+,-,n} {\bf \tilde{1}}_{n_m\times n_m}   & 0 & V {\bf 1}_{n_m\times n_m}
 \end{array}\right)\] 
An important point to note is that off diagonal term appears when the Zeeman effect is important. Such terms are absent in graphene and are responsible for g-factor dependent contributions to transport. 
By performing analytical continuation ($i\omega\rightarrow\omega+i0^+$), the diagonal and off diagonal 
Green's functions (the off diagonal part describes the change from $\gamma$ to $-\gamma$ state or particle hole transition) are
\begin{widetext}
\begin{eqnarray}\nonumber
&&G(k,n,\gamma;\omega+0^+)=\frac{\omega+\mu+\gamma\epsilon_n-\Sigma(\gamma,n,\omega)}{(\omega+\mu-\Sigma(\gamma,n,\omega)-\Sigma(\gamma,-\gamma,n,\omega))(\omega+\mu-\Sigma(\gamma,n,\omega)+\Sigma(\gamma,-\gamma,n,\omega))-\epsilon_n^2}\\\label{gr}
&&G(k,0;\omega+0^+)=\frac{1}{\omega+\mu-\epsilon_0-\Sigma(0,\omega)}\\\nonumber
&&G(k,n,\gamma,-\gamma;\omega+0^+)=\frac{\Sigma(\gamma,-\gamma,n,\omega)}{(\omega+\mu-\Sigma(\gamma,n,\omega)-\Sigma(\gamma,-\gamma,n,\omega))(\omega+\mu-\Sigma(\gamma,n,\omega)+\Sigma(\gamma,-\gamma,n,\omega))-\epsilon_n^2}
\end{eqnarray}
\end{widetext}
with diagonal self energy $\Sigma(\gamma,n,\omega)$ at Landau level $n$, self energy $\Sigma(0,\omega)$ at zeroth Landau level, and off diagonal self energy $\Sigma(\gamma,-\gamma,n,\omega)$ given by
\begin{eqnarray}\label{selfenergy1}
&&\Sigma(\gamma,n,\omega)=Vn_i/(1-V Z(\gamma,n,\omega))\\\nonumber
&&\Sigma(0,\omega)=Vn_i/(1- V Z(0,\omega))\\\nonumber
&&\Sigma(\gamma,-\gamma,n,\omega)=\frac{n_i V_{\gamma,-\gamma,n}}{1-V_{\gamma,-\gamma,n}Z(\gamma,-\gamma,n,\omega)}
\end{eqnarray}
The explicit expression for internal propagators, $Z(\gamma,n,\omega)$, $Z(0,\omega)$, and $Z(\gamma,-\gamma,n,\omega)$ are given in the Appendix~\ref{A2}. Within these internal propagator
the sum over $k$ (in $y$ direction) gives a factor $g_c=\bar{V}/{2\pi l_B^2}=2/(k_c^2l_B^2)=2v_F^2/(D^2l_B^2)$, which accounts for degeneracy of a given Landau level per unit cell and we drop the $k$ dependence in the Green function in the Appendix~\ref{A2}. 
The self consistent solutions of Eq.(\ref{gr}) and Eq.(\ref{selfenergy1}) gives the density of states and the renormalization of Landau levels due to impurities\cite{Peres}.
In the case of weak magnetic field where the Zeeman contribution can be neglected we obtain the same results, modulo a constant factor due to spin and valley degeneracies in graphene, as in Ref.~\onlinecite{Peres}.

 Our main goal is to study the transport properties of surface states. 
The particle current current correlation $L^{11}_{xx}(\omega)$ and $L^{11}_{xy}(\omega)$ from Eq.(\ref{kubo}) are given in the Appendix~\ref{A3}.  
Before we proceed to evaluate the conductivity $\sigma_{ij}$ and thermal power $S_{ij}$ numerically for general impurity strength, we take the clean system limit $V\rightarrow 0$ and study analytically the thermopower in the two dimensional helical metal. We show that in this clean limit we obtain universal feature for diagonal thermopower, similar to its counterpart in the conventional two dimensional electron gas. 
\subsection{Analytic results for weak impurities}
For weak impurities $V\simeq 0$ we have $\Im G(k,n;\epsilon)\simeq \pi\delta(\epsilon+\mu-\epsilon_n)$. Taking this expression into Eq.(\ref{l11xx}) gives 
\begin{eqnarray*}
L^{11}_{xx}|_{\omega\rightarrow 0}=L^{11}_{yy}|_{\omega\rightarrow 0}=0
\end{eqnarray*}
since the energy difference between Landau levels is large enough to prevent the transitions between different levels without the help of impurity broadening.
Also from Eq.(\ref{l11xy}) $L^{11}_{xy}=-L^{11}_{yx}$ with $L^{11}_{xy}$ given by
\begin{eqnarray}\nonumber
&L_{xy}^{11}&=\frac{-T v_F^2}{4\pi l_B^2 \omega\hbar}\int \frac{d\epsilon}{2\pi} \tanh(\frac{\epsilon}{2T})\sum_{\alpha,\gamma}\Big[\frac{\gamma \pi\delta(\epsilon+\mu-\epsilon_0)}{\epsilon+\mu+\gamma\omega-\alpha\epsilon_1}\\\nonumber
&-&\frac{\gamma \pi\delta(\epsilon+\mu-\alpha\epsilon_1)}{\epsilon+\mu+\gamma\omega-\epsilon_0}+\sum_{\lambda,n}\Big(\frac{\pi\delta(\epsilon+\mu-\lambda\epsilon_n)\gamma/2}{\epsilon+\mu+\gamma\omega-\alpha\epsilon_{n+1}}\\\nonumber
&-&\frac{\pi\delta(\epsilon+\mu-\lambda\epsilon_{n+1})\gamma/2}{\epsilon+\mu+\gamma\omega-\alpha\epsilon_n}\Big)\Big]\\\nonumber
&=&\frac{-T v_F^2}{4\pi l_B^2 \omega\hbar}\Big\{\tanh(\frac{\epsilon_0-\mu}{2T})\Big[\frac{\epsilon_0+\omega}{(\epsilon_0+\omega)^2-\epsilon_1^2}\\\nonumber&-&\frac{\epsilon_0-\omega}{(\epsilon_0-\omega)^2-\epsilon_1^2}\Big]
-\frac{\tanh(\frac{\epsilon_1-\mu}{2T})\omega}{(\epsilon_1-\epsilon_0)^2-\omega^2}-\frac{\tanh(\frac{-\epsilon_1-\mu}{2T})\omega}{(\epsilon_1+\epsilon_0)^2-\omega^2}\\\nonumber
&+&\sum_n \frac{1}{2}\Big(
(\tanh(\frac{\epsilon_n-\mu}{2T})+\tanh(\frac{-\epsilon_n-\mu}{2T}))\\\nonumber&\times&\Big[\frac{\epsilon_n+\omega}{(\epsilon_n+\omega)^2-\epsilon_{n+1}^2}-\frac{\epsilon_n-\omega}{(\epsilon_n-\omega)^2-\epsilon_{n+1}^2}\Big]
\\\nonumber
&-&(\tanh(\frac{\epsilon_{n+1}-\mu}{2T})+\tanh(\frac{-\epsilon_{n+1}-\mu}{2T}))\\\label{eq11}
&\times&\Big[\frac{\epsilon_{n+1}+\omega}{(\epsilon_{n+1}+\omega)^2-\epsilon_{n}^2}-\frac{\epsilon_{n+1}-\omega}{(\epsilon_{n+1}-\omega)^2-\epsilon_{n}^2}\Big]\Big)\Big\}
\end{eqnarray}
From Eq.(\ref{eq11}) it is easy to see that for $\epsilon_0=\mu=0$ we have $L_{xy}^{11}=0$ as a result of particle hole symmetry.
For $L_{xy}^{11}|_{\omega\rightarrow 0}$ we have
\begin{eqnarray}\nonumber
&&L_{xy}^{11}|_{\omega\rightarrow 0}=\frac{-T}{8\pi\hbar}\Big\{-2\tanh(\frac{\epsilon_0-\mu}{2T})-\frac{\omega_c^2\tanh(\frac{\epsilon_1-\mu}{2T})}{(\epsilon_1-\epsilon_0)^2}\\\nonumber
&&+\frac{\omega_c^2\tanh(\frac{\epsilon_1+\mu}{2T})}{(\epsilon_1+\epsilon_0)^2}-\sum_{n=1}^{n_{max}} [(\tanh(\frac{\epsilon_n-\mu}{2T})-\tanh(\frac{\epsilon_n+\mu}{2T}))\\\label{eq27}
&&+(\tanh(\frac{\epsilon_{n+1}-\mu}{2T})-\tanh(\frac{\epsilon_{n+1}+\mu}{2T}))]\Big\}
\end{eqnarray}
Here $\omega_c=\sqrt{2}v_F/l_B$ is the cyclotron frequency.
 Eq.(\ref{eq27}) suggests that the hall conductivity can be used as a probe for nonzero $\epsilon_0$, or in other words, whether the Dirac spectrum is gapped by the Zeeman term under the assumption that the sample is clean. By using Eq.(\ref{eq27}) in Eq.(\ref{l12}) we get
\begin{eqnarray*}
&&L_{xy}^{12}|_{\omega\rightarrow 0}=\frac{-T}{8\pi\hbar}\int_{-\infty}^{\infty}d\epsilon (\epsilon-\mu) \left(-\frac{\partial n_F(\epsilon-\mu)}{\partial\epsilon}\right)\\&&\Big\{[\Theta(\epsilon_0-\epsilon)
-\Theta(\epsilon-\epsilon_0)](-2)-(\frac{\omega_c^2}{(\epsilon_1-\epsilon_0)^2}+1)[\Theta(\epsilon_1-\epsilon)\\&&-\Theta(\epsilon-\epsilon_1)]
+(\frac{\omega_c^2}{(\epsilon_1+\epsilon_0)^2}+1)[\Theta(\epsilon+\epsilon_1)-\Theta(-\epsilon_1-\epsilon)]\\
&&-2\sum_{n=2}^{n_{max}}[(\Theta(\epsilon_n-\epsilon)-\Theta(\epsilon-\epsilon_n))-(\Theta(\epsilon_n+\epsilon)\\&&-\Theta(-\epsilon-\epsilon_n))]
\Big\}
\end{eqnarray*}

In the clean limit the off diagonal thermopower $S_{xy}=S_{yx}=0$. The diagonal component $S_{xx}=S_{yy}$, in the low temperature limit (i.e. $\hbar\omega_c\gg k_B T$) with chemical potential close to Landau level $\epsilon_n$, is given by
\begin{equation}\label{clean}
S_{xx}(\mu,T\rightarrow 0)\simeq 
\left\{ \begin{array}{rl}
 &\frac{4\ln2\frac{k_B}{e}}{(\frac{\omega_c^2}{(\epsilon_1-\epsilon_0)^2}-\frac{\omega_c^2}{(\epsilon_1+\epsilon_0)^2})}  \mbox{ if $\mu=\epsilon_0$} \\
 & \frac{\mp 2\ln2\frac{k_B}{e}(\frac{\omega_c^2}{(\epsilon_1\mp\epsilon_0)^2}+1)}{3+(\frac{\omega_c^2}{(\epsilon_1\pm\epsilon_0)^2})}  \mbox{ if $\mu=\pm\epsilon_1$} \\
 & \frac{\mp 4\ln2\frac{k_B}{e}}{4N-2+(\frac{\omega_c^2}{(\epsilon_1-\epsilon_0)^2}+\frac{\omega_c^2}{(\epsilon_1+\epsilon_0)^2})}\\& \mbox{if $\mu=\pm\epsilon_N$ and $N\ge 2$}
       \end{array} \right.
\end{equation}
\begin{figure}[t]
\includegraphics[width=1\columnwidth, clip]{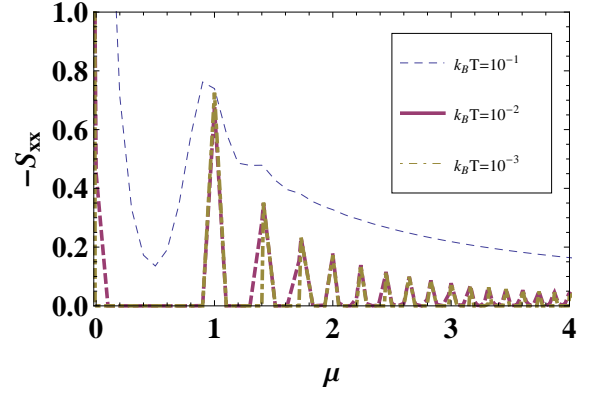}
\caption{Negative thermopower $-S_{xx}$ as a function of chemical potential $\mu$ for temperature $k_B T=10^{-1}$ (Blue, Dashed), $10^{-2}$ (Purple, Thick line), and $10^{-3}$ (Brown, Dot Dashed) in the clean helical metal. $\mu$ and $k_BT$ are in unit of $\hbar\omega_c$ and $S_{xx}$ is in unit of $k_B/e$. Around $\mu=0$ $-S_{xx}$ becomes order of $10^1\frac{k_B}{e}$ for all three cases. The peaks value are in nice agreement with Eq.(\ref{clean}).}
	  \label{fig0}
\end{figure}
\begin{figure}[t]
\includegraphics[width=1\columnwidth, clip]{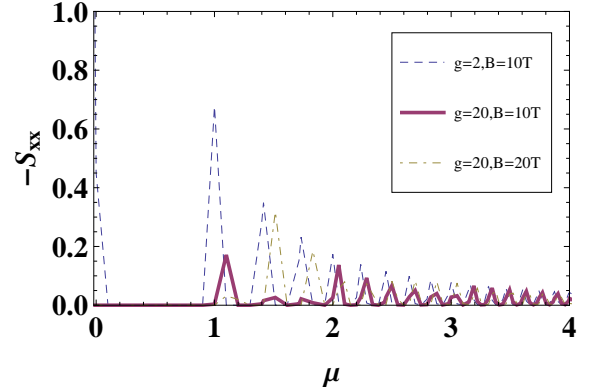}
\caption{Negative thermopower $-S_{xx}$ as a function of chemical potential $\mu$ for temperature $k_B T=10^{-2}\hbar\omega_c$ with gyromagnetic ratio $g=2$ (blue thin line), $g=20$ and $B=10T$ (purple thick line), and $g=20$ and $B=20T$ (brown dot dashed line) in the clean helical metal. $\mu$ and $k_BT$ are in unit of $\hbar\omega_c$ and $S_{xx}$ is in unit of $k_B/e$. The peak height is not universal with large Zeeman effect.}
	  \label{fig01}
\end{figure}
We can approximate $(\epsilon_1\mp\epsilon_0)^2\simeq\epsilon_1^2(1\mp 2\frac{\epsilon_0}{\epsilon_1})$ because $\alpha(\simeq 10^{-4}(1/T))$ is small. Under this approximation we see that for $\mu=\epsilon_0=g\mu_B B\sin\theta$ the thermopower becomes $-\frac{k_B}{e}\ln 2 \frac{\epsilon_1}{\epsilon_0}$, which again serves as a measurable quantity for the Zeeman term. In real material, however, this singular behavior around $\epsilon_0$ will be masked by the broadening of zeroth Landau level due to impurities or Coulomb interaction. For materials with negligible Zeeman interaction the thermopower reaches $\frac{\mp k_B\ln2}{N e}$  for $\mu=\pm\epsilon_N$, irrespective of the strength of magnetic field. This is consistent with the Berry phase argument\cite{Zuev,Check}. For weak magnetic field strength ($\epsilon_0\ll \epsilon_1$) the thermopower of two dimensional helical metal is similar to that of the conventional 2D metal in that there exists universality for peak value of thermopower in the clean limit\cite{Jonson}. The feature observed in the helical metal, however, is different from the case of conventional 2D metal where every Landau level $N$ has universal value $-\frac{k_B\ln2}{e(N+\frac{1}{2})}$. 

From Eq.(\ref{clean}) we see that for large Zeeman effect ($\epsilon_1-\epsilon_0>\omega_c$) the peak height of $S_{xx}$ is no longer universal and depends on 
the strength of Zeeman interaction. We artificially increases this Zeeman effect by increasing the gyromagnetic ratio and magnetic field strength. The results are shown in Fig.\ref{fig01}. In addition to the non-universality another feature to note is the shift in the position of the peaks reflecting the Zeeman contribution to the energy. Such a shift is most prominent for the lowest Landau levels, increasing the spacing for positive $N$. The significant reduction in the peak values as a function of magnetic field, as compared to graphene, provides a clear signature of the surface state. One consequence is that for the same impurity broadening, systems with larger Zeeman coupling possess gaps in the spectrum near the zeroeth Landau level even with the broadening from impurity interaction. The effect of impurities are explored in the next subsection.
 
\subsection{General impurity strength}
For general impurity strength and large magnetic field strength we expand $f_{\gamma,n}\simeq(1+\frac{\alpha B\sin\theta}{4n}+\frac{g\mu_B l_B B\sin\theta}{\gamma v_F\sqrt{2n}})$. Within this approximation $V_{\gamma,-\gamma,n}=\frac{-V\alpha B\sin\theta}{4n}$.  The off-diagonal self energy $\Sigma(\gamma,-\gamma,n,\omega)$ is
\begin{eqnarray}\label{rov}
\Sigma(\gamma,-\gamma,n,\omega)=\frac{n_i\frac{-V\alpha B\sin\theta}{4n}}{1-\frac{g_c V\alpha B\sin\theta}{8n}\frac{1}{\omega+\mu-\epsilon_0-\Sigma(0,\omega)}}
\end{eqnarray}
 The $4n_{max}+1$ self energy terms in self consistency equations Eq.(\ref{gr}) and Eq.(\ref{selfenergy1}) are therefore simplified to two variables $\Sigma(0,\omega)\equiv\Sigma_0(\omega)$ and $\Sigma(\gamma,n,\omega)\equiv\Sigma(\omega)$ which we evaluated numerically in this section. 
  We put band energy $D=3.5eV$, Fermi velocity $v_F=5\times 10^5 m/s$, and impurity density $n_{imp}=10^{-3}$ in the following numerical computations.
  
\begin{figure}[t]
\includegraphics[width=.5\columnwidth, clip]{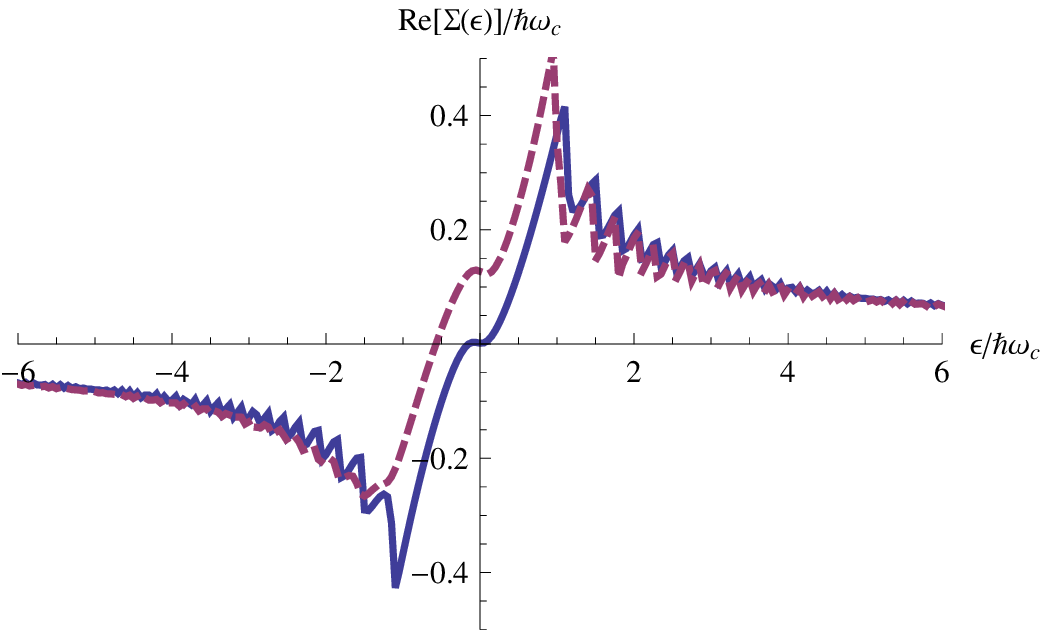}\hfill
\includegraphics[width=.5\columnwidth, clip]{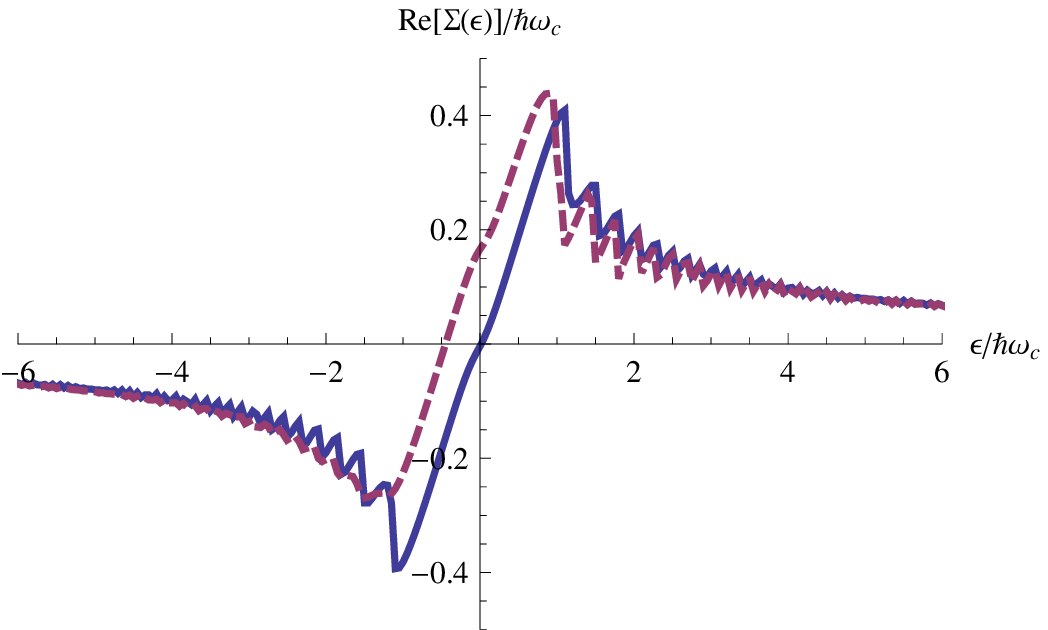}\\
\includegraphics[width=.5\columnwidth, clip]{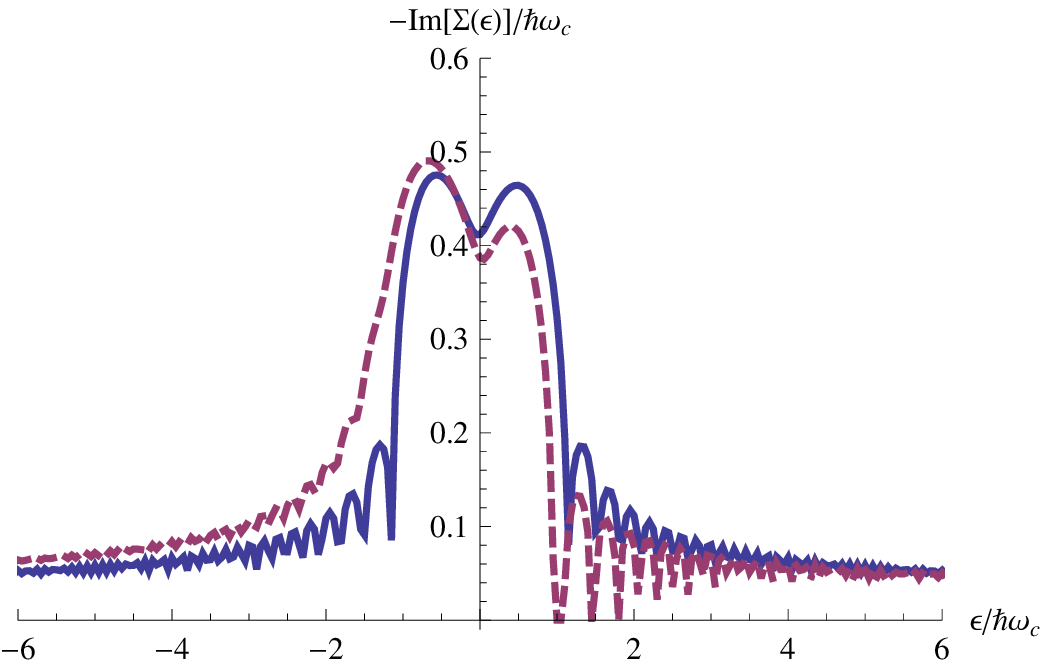}\hfill
\includegraphics[width=.5\columnwidth, clip]{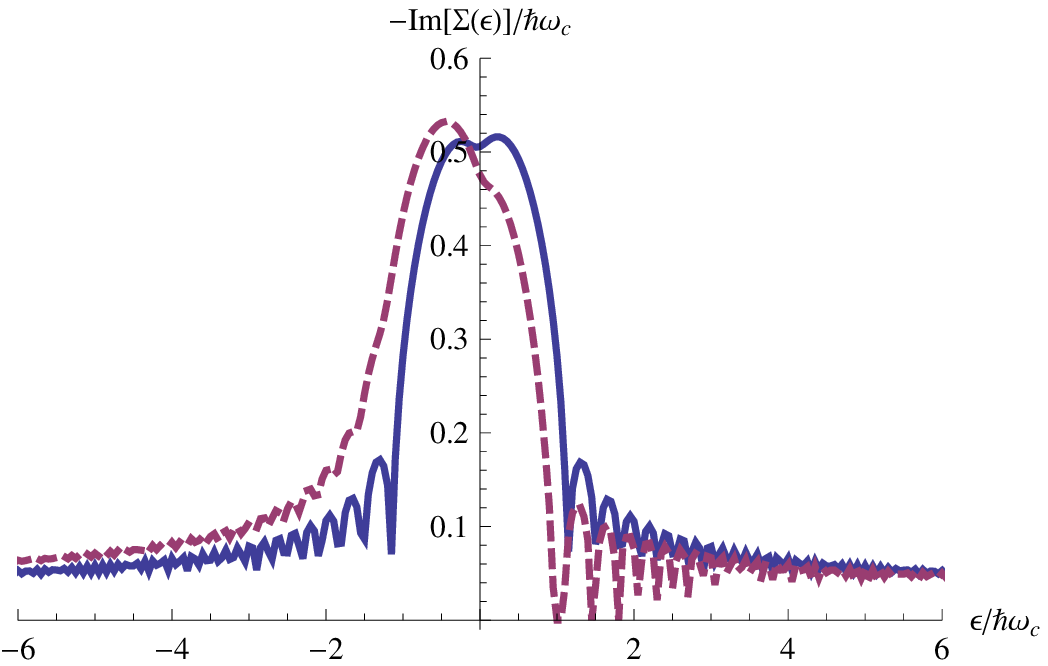}
\caption{Self energy $\Sigma_0(\epsilon)$ (left) and $\Sigma(\epsilon)$ (right) vs $\epsilon$. Energy is in unit of $\hbar\omega_c$ and $B=14T$ in all figures. Top panel shows real part and bottom panel shows imaginary part. Impurity strength $V=10^5\hbar\omega_c$ (blue lines), $V=10^3\hbar\omega_c$ (purple dashed lines), and impurity concentration $n_i=10^{-3}$.}
	  \label{fig2}
\end{figure}

The real and imaginary part of the self energy are shown in Fig.\ref{fig2} for $B=14T$. For strong impurity strength (blue lines) the imaginary part of the self energy shows particle hole symmetry ($\mu=0$ in all self energy figures) while for moderate impurity strength (purple dashed lines) the self energy shows more weight on the hole side. In clean limit, as discussed in last section, or with weak impurity strength the particle hole symmetry of the self energy is restored. This can be understood by taking $V\rightarrow\infty$ for vacancies and $V\rightarrow 0$ for weak impurity in Eq.(\ref{SCBA}). For $V\rightarrow\infty$, $\Sigma_0\mathbb{G}\gg {\bf 1}$ and thus $\Sigma=-n_i\mathbb{G}^{-1}$, which renders the self energy to be particle hole symmetric. For $V\rightarrow 0$, $\Sigma=n_i\Sigma_0({\bf 1}-\Sigma_0\mathbb{G})^{-1}\simeq n_i\Sigma_0+ n_i\Sigma_0^2\mathbb{G}$. The first constant term can be absorbed in the shift of chemical potential and therefore the self energy obtained by second order self consistency equation is particle hole symmetric. 
For intermediate impurity strength, the factor ${\bf 1}$ in numerator of Eq.(\ref{SCBA}) is comparable with $\Sigma_0\mathbb{G}$. Thus for repulsive $V$ the self energy is shifted to the left.

\begin{figure}[t]
\includegraphics[width=.5\columnwidth, clip]{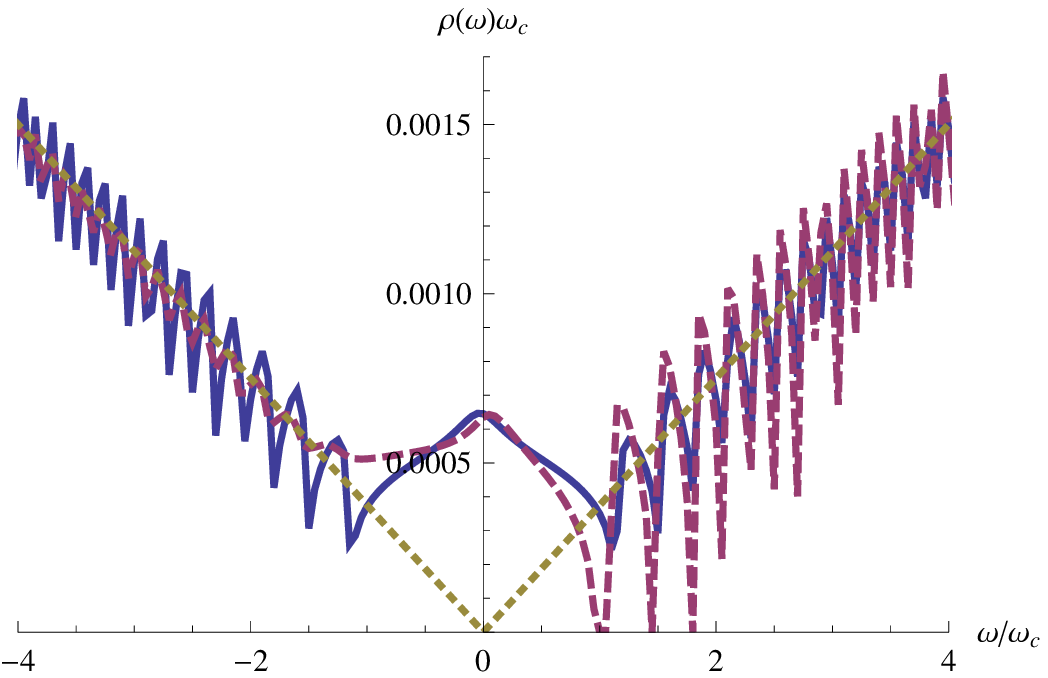}\hfill
\includegraphics[width=.5\columnwidth, clip]{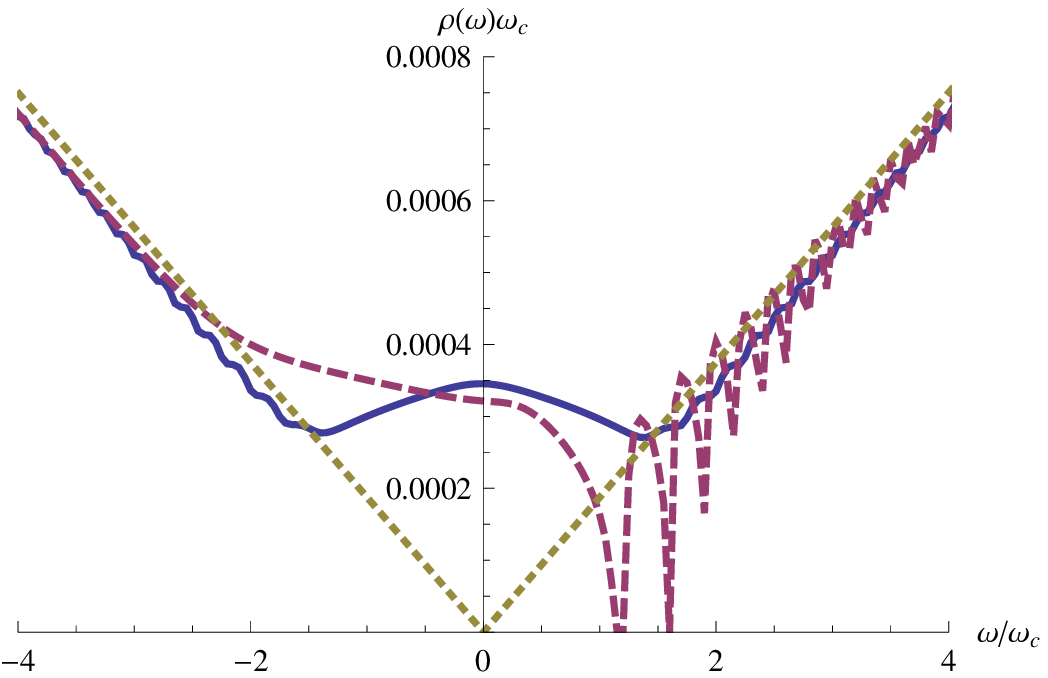}
\caption{Density of state per unit cell in unit of $1/\hbar\omega_c$ vs $\omega/\omega_c$ for $B=14T$ (left) and $B=7T$ (right). Blue lines are $V=10^5\hbar\omega_c$ and purple dashed lines are for $V=10^3\hbar\omega_c$. The brown dotted lines indicate the density of state with degeneracy factor $g_c$ and without impurities.}
	  \label{fig2b}
\end{figure}

\begin{figure}[t]
\includegraphics[width=.5\columnwidth, clip]{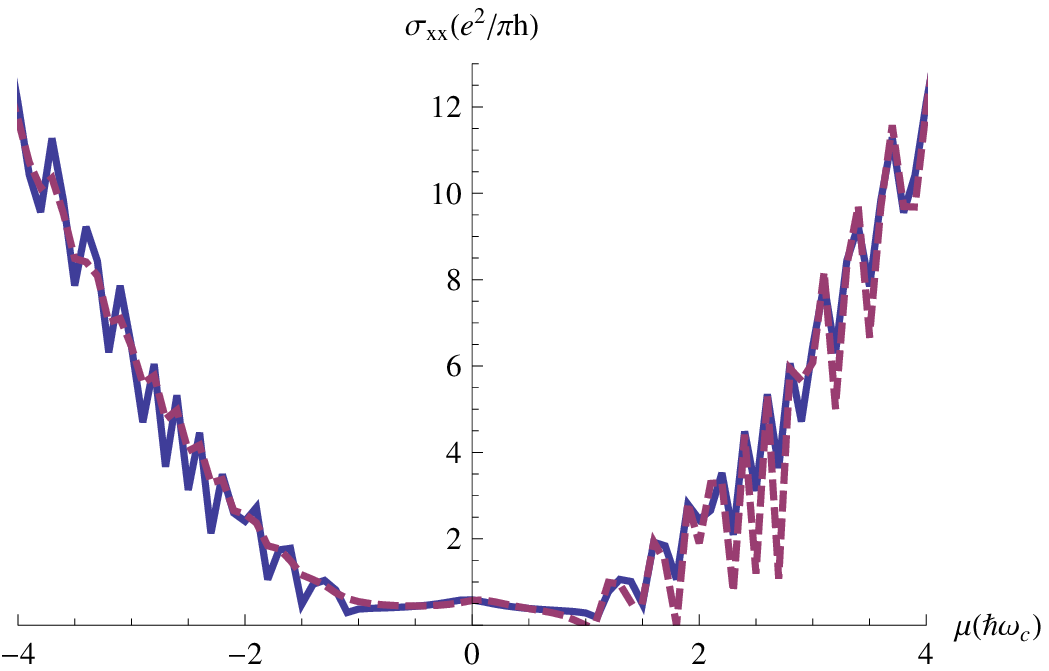}\hfill
\includegraphics[width=.5\columnwidth, clip]{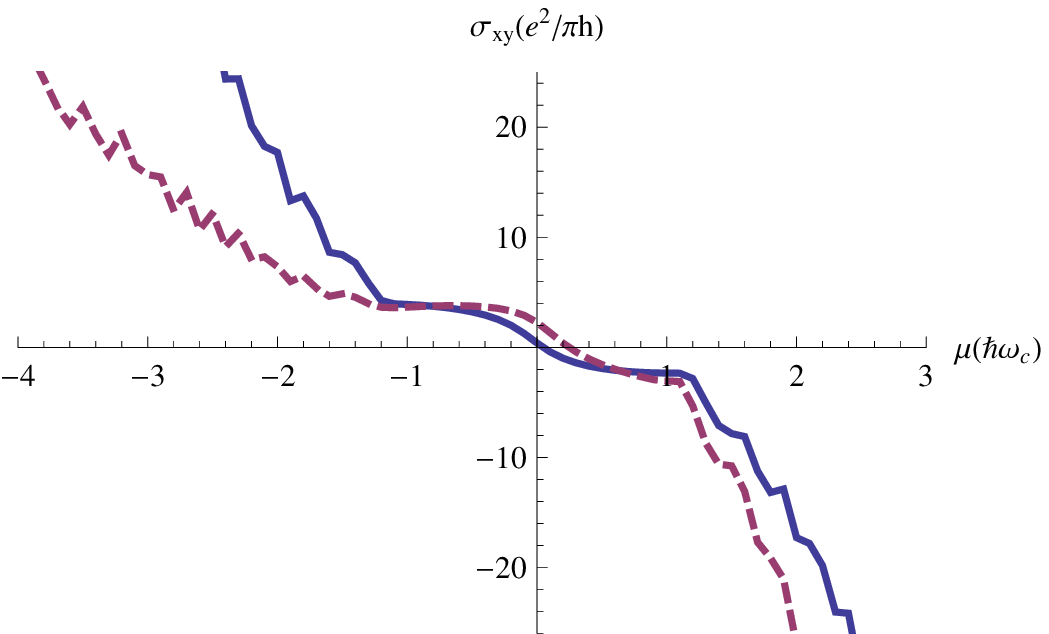}\\
\includegraphics[width=.5\columnwidth, clip]{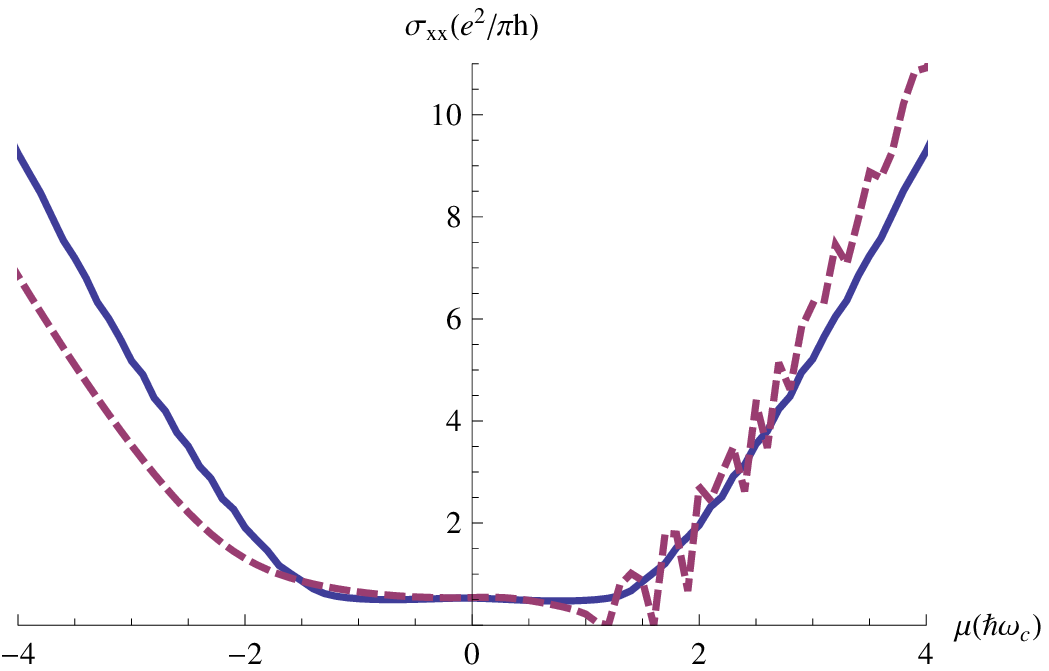}\hfill
\includegraphics[width=.5\columnwidth, clip]{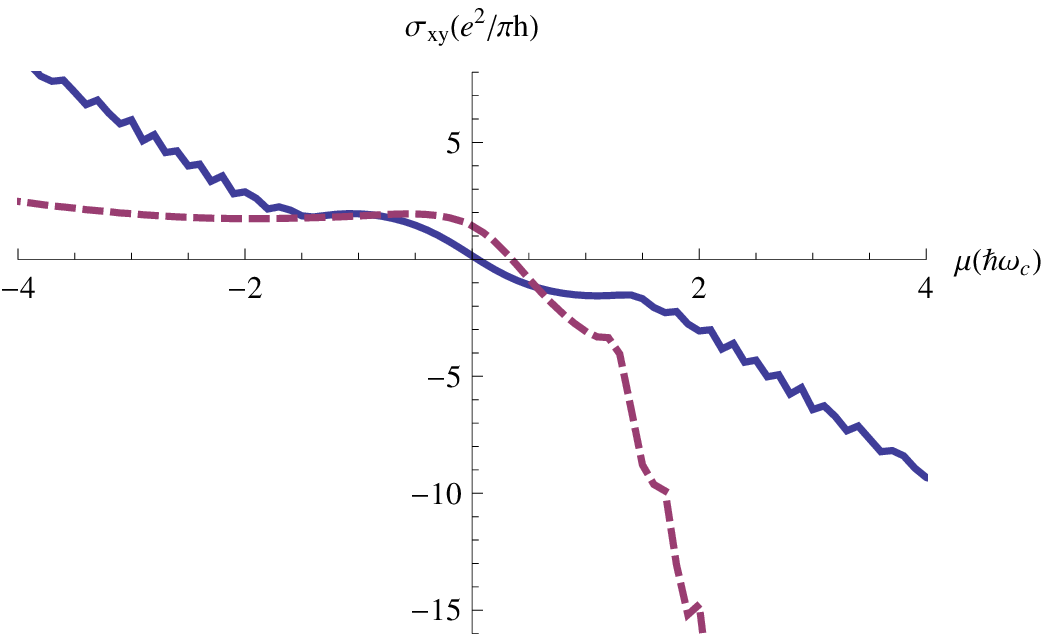}
\caption{DC conductivity $\sigma_{xx}$ (left) and $\sigma_{xy}$ (right) as a function of chemical potential $\mu$. Conductivity is in unit of $e^2/h\pi$. $\mu$, $k_BT=10^{-6}$, impurity strength $V=10^5$ (blue lines), and $V=10^3$ (purple dashed lines) are in unit of $\hbar\omega_c$. Magnetic field strength $B=14T$ for upper and $B=7T$ for lower figures. The inset of upper right figure shows the blowup plot with smaller chemical potential range. The $\sigma_{xy}$ in $B=14T$, $V=10^3\hbar\omega_c$ case has been shifted down by $50e^2/h$ and $\sigma_{xy}$ in $B=7T$, $V=10^3\hbar\omega_c$ case has been shifted down by $3.3e^2/h$ to match the correct charge neutral point.}
	  \label{fig3}
\end{figure}

From the self energy we obtain the density of state per unit cell as
\begin{eqnarray}
\rho(\omega)=\frac{-g_c}{\pi}\Im[G(0;\omega+0^+)+\sum_{\gamma,n}G(n,\gamma;\omega+0^+)]
\end{eqnarray}
The numerical results are shown in Fig.\ref{fig2b} for two different magnetic field strengths. For $V=10^3\hbar\omega_c$ we see clear signature of particle hole asymmetry and there are more holes than 
electrons when $\mu=0$. The Hall conductivity $\sigma_{xy}$ obtained in this case does not flip its signs at $\mu=0$. Rather the charge neutrality point has been shifted to positive $\mu$. Another feature is that the oscillation due to applied magnetic field is more apparent in $\omega>0$ region than in $\omega<0$ region for $V=10^3\hbar\omega_c$. This feature also shows up in the longitudinal conductivity $\sigma_{xx}$ v.s. chemical potential. The numerical results for DC conductivity $\sigma_{xx}$ and $\sigma_{xy}$ as a function of chemical potential $\mu$ are plotted in Fig.\ref{fig3}. As is expected from the results of self energy and density of state, both $\sigma_{xx}$ and $\sigma_{xy}$ show more oscillatory behavior/larger steps for $\mu>0$ region.  

\begin{figure*}[tb]
\includegraphics[width=.45\textwidth, height=.2\textwidth, clip]{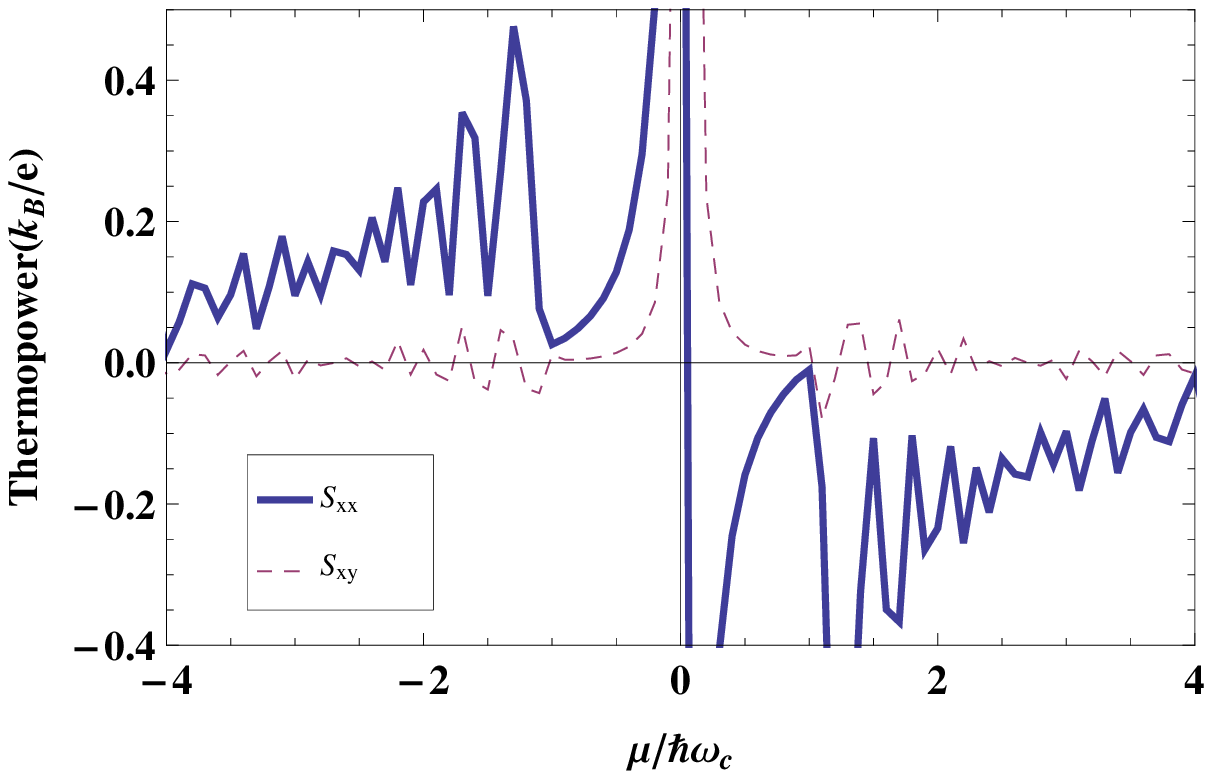}\hfill
\includegraphics[width=.45\textwidth, height=.2\textwidth, clip]{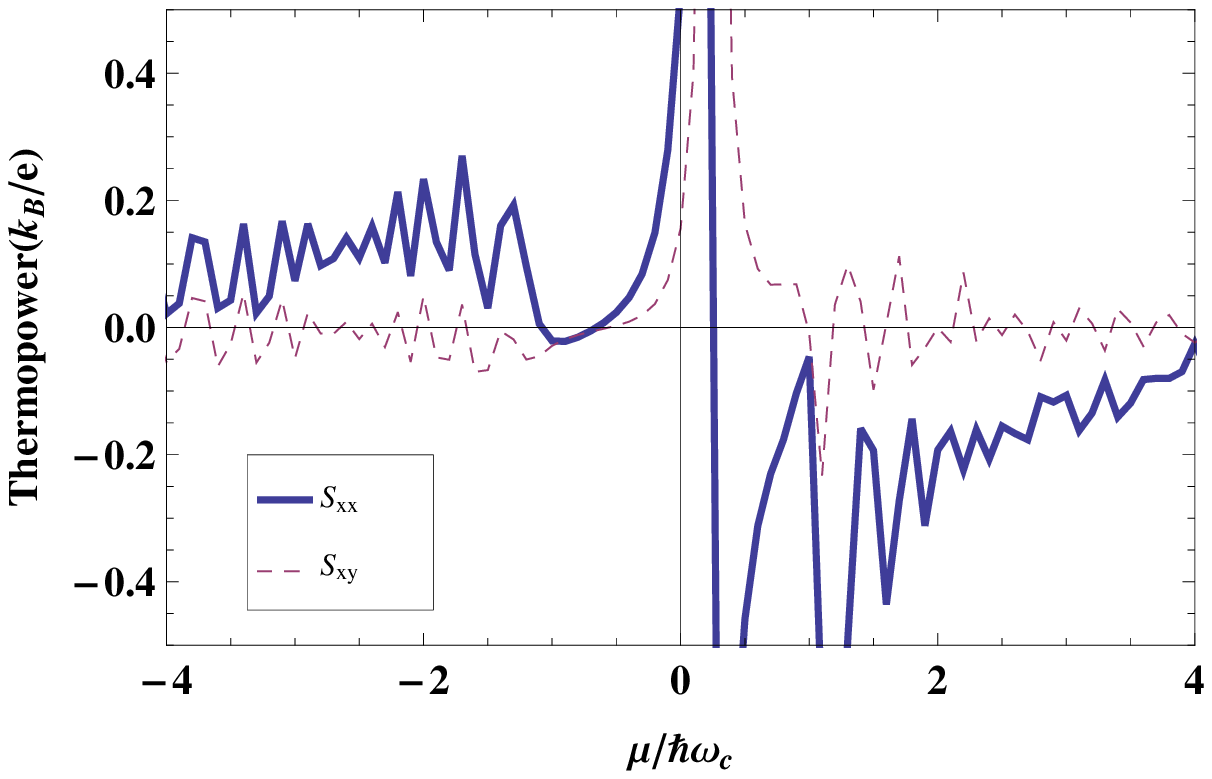}\\
\includegraphics[width=.45\textwidth, height=.2\textwidth, clip]{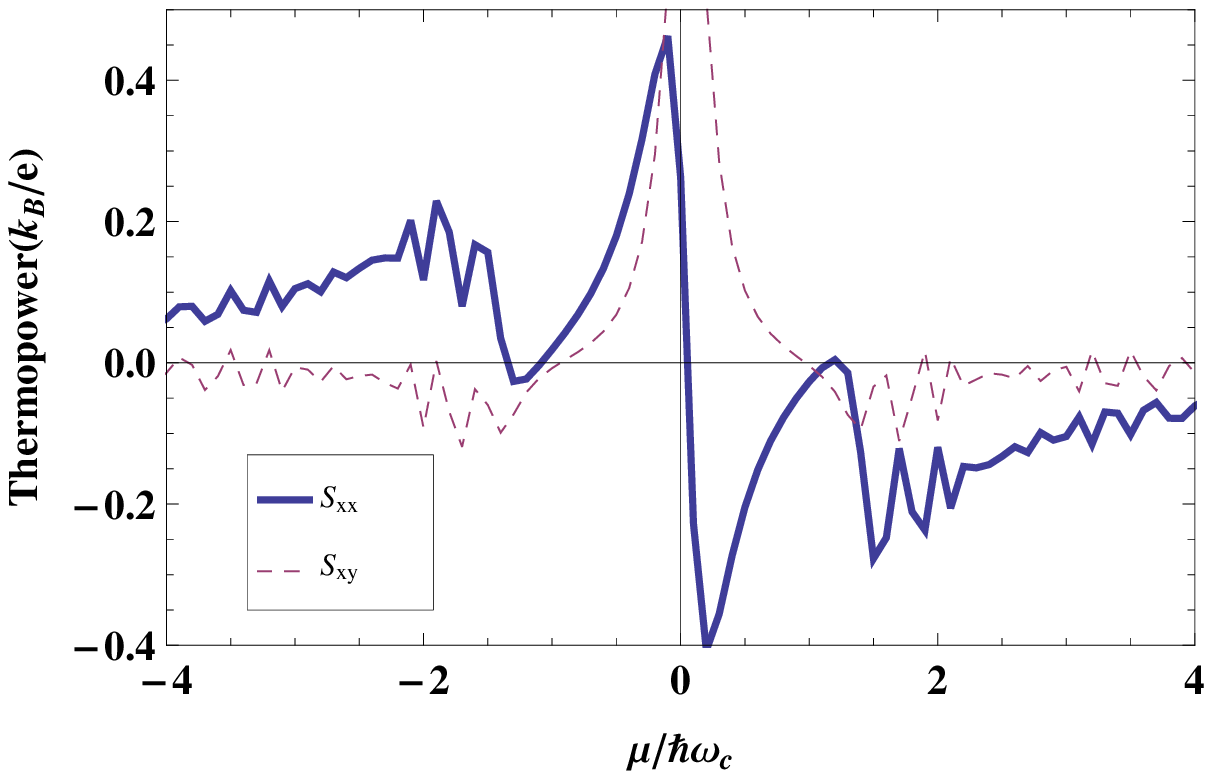}\hfill
\includegraphics[width=.45\textwidth, height=.2\textwidth, clip]{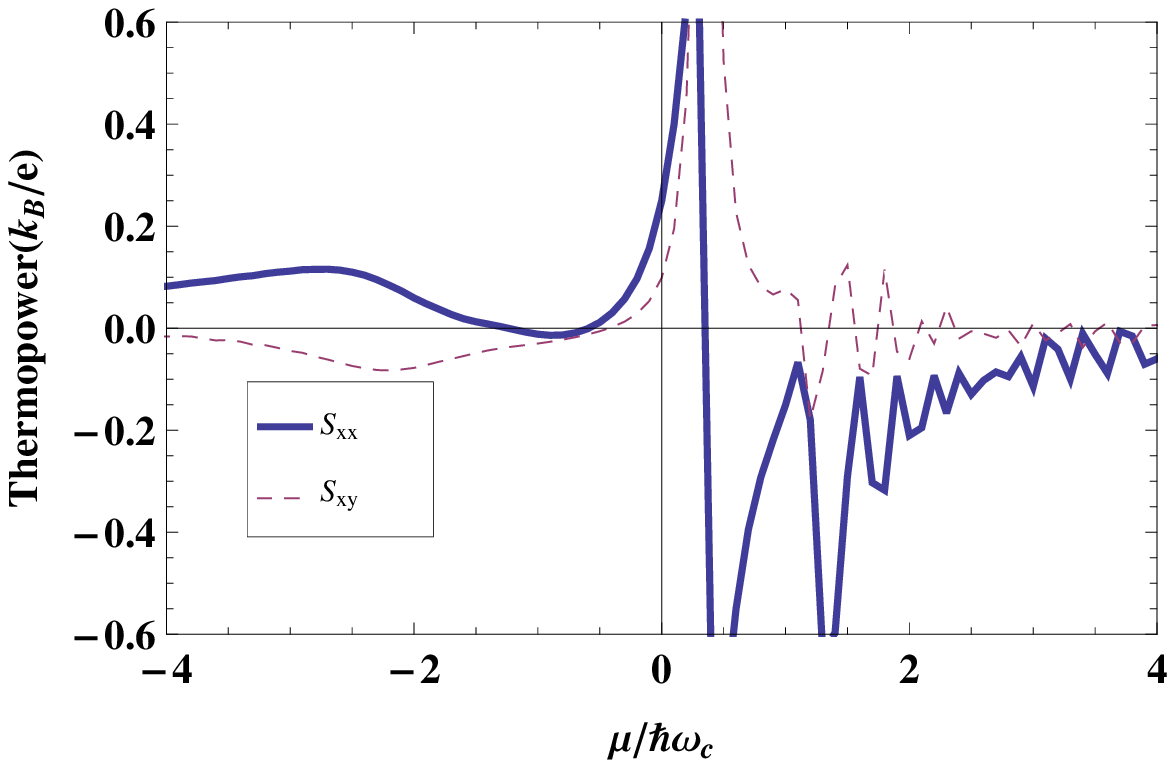}
\caption{Thermopower $S_{xx}$ and $S_{xy}$ as a function of chemical potential $\mu$. The thermal power is in unit of $k_B/e$ and $\mu$, $k_BT$, impurity strength $V$ are in unit of $\hbar\omega_c$. Magnetic field strength $B=14$ for upper two and $B=7T$ for lower two figures. The two left figures show the case of $V=10^5\hbar\omega_c$ and two right figures show
the case of $V=10^3\hbar\omega_c$. $k_BT=0.05\hbar\omega_c$ in all cases.}
	  \label{fig4}
\end{figure*}

 The thermopower $S_{xx}$ and $S_{xy}$ obtained from Eq.(\ref{tp}) are shown in Fig.\ref{fig4} for $B=14T$ and $B=7T$ with $V=10^5\hbar\omega_c$ and $V=10^3\hbar\omega_c$. The general features for $S_{xx}$ are the peak positions corresponding to different Landau levels are shifted away from the $n=0$ Landau level and the width of the peaks increases due to the impurity broadening, similar to effect of temperature. There is no universal value at the peaks of $S_{xx}$ due to impurities as in the case of conventional two dimensional electron gas\cite{Jonson}. Near the charge neutrality point $S_{xy}$ shows a peak while $S_{xx}$ shows singular behavior around this point. The oscillatory behavior in $S_{xx}$ and $S_{xy}$ is more apparent for larger magnetic field, while the asymmetry is more prominent at lower fields. Note that the oscillations are entirely absent for negative chemical potentials for moderate impurity potentials. These results are in qualitative agreement with available scanning tunneling microscopy(STM) data which find oscillations in density of states only for positive gate voltages\cite{Hanaguri, Pchen}.
 
Currently only limited experimental results for the thermopower of helical metals are available. Since our formulation is very similar to the case of graphene, we compare our thermopower results with data on graphene. In Ref.~\onlinecite{Zuev} the observed peaks of magneto-thermopower $S_{xx}$ shows the quantized $1/N$ ($N$ is the nonzero Landau level index) trend but with a reduced factor in height and peak broadening, which are qualitatively consistent with our numerically results shown in Fig.\ref{fig4} for $V=10^5\hbar\omega_c$. In Ref.~\onlinecite{Check} similar results have been reported, with the measured peak value of $S_{xx}$ at $N=-1$ and $B=9T$ is $41\mu V/K$, which is smaller than the clean limit where $S_{xx}=\ln2 k_B/e=0.69 k_B/e \simeq 59.7\mu V/K$. The thermopower shown in Fig.\ref{fig4} for $B=14T$, $V=10^5\hbar\omega_c$ is similar to the case seen in Ref.~\onlinecite{Check} that $S_{xx}\simeq 40\mu V/K$ for $N=-1$ and smaller $S_{xx}$ at $|N|>1$, suggesting the influence of impurity interactions. In both experiments a large peak at $N=0$ Landau level is observed which is consistent with our numerical results. This peak in $S_{xy}$ can be explained by the presence of impurities in addition to the Dirac spectrum.

The major difference between our results and experimental results in graphene\cite{Zuev,Check,Wei} is the behavior of $S_{xx}$ near $N=0$ region. Our numerical results suggest a large electron/hole peak between $N=0$ and $N=1/-1$, showing remnant singular behavior around $N=0$ landau level in the clean limit as seen in Fig.\ref{fig0}. In experimental data a hole/electron peak between $N=0$ and $N=1/-1$ is observed instead. This discrepancy is related to the higher $\sigma_{xx}$ value when $\mu$ is at $N=0$ Landau level, as confirmed by the following observations. In all our numerical results the diagonal conductivity $\sigma_{xx}$ is close to the universal value $e^2/2\pi h$ at $\mu=0$. By adding a Lorentzian profile to artificially increase $\sigma_{xx}$ around $\mu=0$ we can reproduce this results around $N=0$ seen in the experiments. Similar observations were reported in Ref.~\onlinecite{Zhu} where they performed numerical study on discrete lattice model for graphene system. The physical reason for the increase of $\sigma_{xx}$ near $N=0$ Landau level is still an open question and we discuss the possible mechanism for this enhancement of $\sigma_{xx}$ near the Dirac node under magnetic field in Sec.~\ref{s5}.  

\section{Conclusion}\label{s5}
In this article we compute linear response conductivity and thermopower of the helical metal, both with and without magnetic field, in the presence of static, short ranged, random impurities. Analytic results of thermopower in the clean limit, i.e. impurity strength $V\rightarrow 0$, are given in both cases. In the case of zero magnetic field the thermopower obtained is still susceptible to impurity interaction and does not show universal feature. In the case of finite magnetic field the thermopower shows universal value, independent of temperature and magnetic field strength, when the chemical potential is fixed around Landau levels at low temperature. We also compute the thermopower and conductivity numerically for the case of finite impurity strength. The effect of impurities broaden the peaks in diagonal thermopower and the height of the peaks show no universal behavior, which is similar to the case in the conventional 2D electron gas\cite{Jonson}. We also vary the impurity potential and study its affect on transport. For moderate impurity strength $V$ we find the particle hole symmetry is broken while for large or small $V$ the symmetry is restored in our perturbation formulation.  

Due to the lack of available data on thermopower measurement of the helical metal, we compare our thermopower results with experimental data on graphene. Our computation is on the helical metal system which differs from graphene\cite{Peres} mainly in the Zeeman effect. Since the Zeeman contribution is small for normal magnetic field strength (due to $\alpha\simeq10^{-4}/T$) in both graphene and helical metal samples such as HgTe, our results are applicable to the case of graphene with short range, randomly positioned impurities modulo a numerical factor in conductivity due to valley and spin degeneracy in graphene. We find qualitative agreement between our numerical results of thermopower and experimental results in graphene, except for the $S_{xx}$ near the zeroth Landau level. The discrepancy is due to the enhanced diagonal conductivity nearby zeroth Landau level. The enhancement in $\sigma_{xx}$ seems to decrease with increasing magnetic field\cite{Check}. Possible mechanism includes the metal to insulator transition on the zeroth Landau level\cite{Zaliznyak} or electron hole puddles \cite{Adam}. Transport measurements of graphene near the minimum conductivity in Ref.~\onlinecite{cho}
partly supports the latter mechanism but further study is needed to clarify this issue. Whether similar behavior is reproduced in helical metals is an open question.

The magneto-electric coupling, unique to topological insulators, leads to novel features in transport. The dependence of the thermopower on the Zeeman coupling provides a diagnostic of the state. To amplify the expected behavior we have analyzed thermopower and Nernst for materials with an artificially enhanced gyromagnetic ratio. The relative importance of the Zeeman term to the Landau level splitting can be enhanced by an in-plane electric field, providing a possible route to detecting the signatures proposed. One caveat the electric field leads to additional effects such as non-linear transport, a detailed analysis of which is the subject of future investigation.

\section*{Acknowledgment}
The authors wish to acknowledge Jing Shi and Peng Wei for useful discussions. Vivek Aji and Sung-Po Chao's research is supported by University of California at Riverside under the initial complement.
\appendix
\section{Derivation of self energy in zero magnetic field case}\label{A0}
The free electron Green's functions in the energy eigenstate basis Eq.(\ref{sp1}) are
\begin{eqnarray*}
&&G_{\uparrow,\uparrow}(\vec{k},\omega_n,)=\sum_{\gamma=\pm 1}\frac{1/2}{i\omega_n+\mu-\gamma v_F k}\\
&&G_{\uparrow,\downarrow}(\vec{k},\omega_n)=\sum_{\gamma=\pm 1}\frac{\gamma e^{i\phi(\vec{k})}/2}{i\omega_n+\mu-\gamma v_F k}\\
&&G_{\downarrow,\uparrow}(\vec{k},\omega_n)=\sum_{\gamma=\pm 1}\frac{\gamma e^{-i\phi(\vec{k})}/2}{i\omega_n+\mu-\gamma v_F k}\\
&&G_{\downarrow,\downarrow}(\vec{k},\omega_n)=G_{\uparrow,\uparrow}(\vec{k},\omega_n)
\end{eqnarray*} 
In this basis, the impurity interaction in Eq.(\ref{hi}) is
\begin{eqnarray*}
H_{imp}=\frac{V}{2 A}\sum_{i=1}^{N_{imp}}\sum_{\vec{p},\vec{k},\gamma,\bar{\gamma}}e^{i(\vec{k}-\vec{p})\cdot\vec{r}}(\gamma\bar{\gamma}+e^{i(\phi(\vec{k})-\phi(\vec{p}))})c_{\vec{k},\gamma}^{\dagger}c_{\vec{p},\bar{\gamma}}
\end{eqnarray*}
The  impurity interaction is diagonal in the spin space. The position averaged Green's function, obtained within SCBA\cite{Bruus, Peres}, is 
\begin{eqnarray*}
&&\langle G_{s,s}(\vec{k},\omega_n;\vec{r}_1,\vec{r}_2,\vec{r}_3,\ldots,\vec{r}_{N_{imp}})\rangle\equiv G(k,\omega_n)\\&&=\frac{1}{A^{N_{imp}}}\Big[\prod_{i=1}^{N_{imp}}\int d\vec{r}_i\Big] G_{s,s}(\vec{k},\omega_n;\vec{r}_1,\vec{r}_2,\vec{r}_3,\ldots,\vec{r}_{N_{imp}})
\end{eqnarray*}
 Thus the bare propagator spin-diagonal component) in Eq.(\ref{tmatrix}) is replaced by full propagator which leads to full self consistent Born approximation\cite{Peres}
\begin{eqnarray}
\Sigma(\omega_n)=\frac{V n_i}{1-\frac{V}{N}\sum_{\vec{k},\gamma}[i\omega_n+\mu-\gamma v_F k-\Sigma(\omega_n)]^{-1}}
\end{eqnarray}
where $\Sigma(\omega_n)$ is the electron self energy. This is independent of momenta due to spherical symmetry in the impurity potential in Eq.(\ref{hi}).
By using analytic continuation $i\omega_n\rightarrow\omega+i0^+$ we have the self energy of electron given by
\begin{eqnarray}
\Sigma(\omega+i0^+)=\frac{V n_i}{1-\frac{V}{N}\sum_{\vec{k},\gamma}[\omega+\mu-\gamma v_F k-\Sigma(\omega)]^{-1}}
\end{eqnarray}
\begin{widetext}
\section{Explicit form of $\kappa(\epsilon,V,n_{imp})$}\label{A1}
The explicit form of $\kappa(\epsilon,V,n_{imp})$ after the momentum integral is given as

\begin{eqnarray}
&&\kappa(\epsilon,V,n_{imp})=\\\nonumber&&-\frac{2 a b D^2 \left(a^2-b^2-D^2\right)+\left(a^2+b^2\right) \left((a-D)^2+b^2\right) \left((a+D)^2+b^2\right) \left(\tan ^{-1}\left(\frac{a-D}{b}\right)+\tan
   ^{-1}\left(\frac{a+D}{b}\right)-2 \tan ^{-1}\left(\frac{a}{b}\right)\right)}{2 a b \left(a^4+2 a^2 (b-D) (b+D)+\left(b^2+D^2\right)^2\right)}
\end{eqnarray}

with $a=\epsilon-\Re\Sigma(\epsilon)$ and $b=\Im\Sigma(\epsilon)$.
\section{Expression for internal propagator terms}\label{A2}
The internal propagators for self energy computation, $Z(\gamma,n,\omega)$, $Z(0,\omega)$, and $Z(\gamma,-\gamma,n,\omega)$ are given by
\begin{eqnarray}\nonumber
&&Z(\gamma,n,\omega)=\sum_{m,\bar{\gamma}}[\frac{(1+f_{\gamma,n}^2f_{\bar{\gamma},m}^2)g_c G(m,\bar{\gamma};\omega+0^+)}{(1+f_{\gamma,n}^2)(1+f_{\bar{\gamma},m}^2)}
+\frac{(1-f_{\gamma,n}^2 f_{\bar{\gamma},m}f_{-\bar{\gamma},m})g_c G(m,\bar{\gamma},-\bar{\gamma};\omega+0^+)}{(1+f_{\gamma,n}^2)\sqrt{(1+f_{\bar{\gamma},m}^2)(1+f_{-\bar{\gamma},m}^2)}}]\\\label{selfenergy1a}
&&+\frac{f_{\gamma,n}^2}{1+f_{\gamma,n}^2}g_c G(0;\omega+0^+)\\\label{selfenergy2a}
&&Z(0,\omega)=\sum_{m,\bar{\gamma}}[\frac{f_{\bar{\gamma},m}^2}{1+f_{\bar{\gamma},m}^2}g_c G(m,\bar{\gamma};\omega+0^+)
-\frac{g_cf_{\bar{\gamma},m}f_{-\bar{\gamma},m}G(m,\bar{\gamma},-\bar{\gamma};\omega+0^+)}{\sqrt{(1+f_{\bar{\gamma},m}^2)(1+f_{-\bar{\gamma},m}^2)}}]+g_c G(0;\omega+0^+)
\\\nonumber
&&Z(\gamma,-\gamma,n,\omega)
=\sum_{\bar{\gamma},m}[\frac{(1-f_{\gamma,n}f_{-\gamma,n}f_{\bar{\gamma},m}^2)g_c G(m,\bar{\gamma};\omega+0^+)}{(1+f_{\bar{\gamma},m}^2)\sqrt{(1+f_{\gamma,n}^2)(1+f_{-\gamma,n}^2)}}
+\frac{(1+f_{\gamma,n}f_{-\gamma,n}f_{\bar{\gamma},m}f_{-\bar{\gamma},m})g_c G(m,\bar{\gamma},-\bar{\gamma};\omega+0^+)}{\sqrt{(1+f_{\gamma,n}^2)(1+f_{-\gamma,n}^2)(1+f_{\bar{\gamma},m}^2)(1+f_{-\bar{\gamma},m}^2)}}]\\\label{selfenergy3a}
&&-\frac{f_{\gamma,n}f_{-\gamma,n}}{\sqrt{(1+f_{\gamma,n}^2)(1+f_{-\gamma,n}^2)}}g_c G(0;\omega+0^+)
\end{eqnarray}
\section{Expression for $L_{xx}^{11}$ and $L_{xy}^{11}$}\label{A3}
For the case of finite magnetic field in the energy eigenstates basis the particle current operators, as in Eq.(\ref{particlecurrent}), are 
\begin{eqnarray}\nonumber
&&J_x=-v_F\Big\{\sum_{k,\gamma}\frac{1}{\sqrt{1+f_{\gamma,1}^2}}(c_{k,0}^{\dagger}c_{k,1,\gamma}+c_{k,1,\gamma}^{\dagger} c_{k,0})
+\sum_{k,n,\gamma,\bar{\gamma}}\frac{\bar{\gamma}f_{\bar{\gamma},n}}{\sqrt{(1+f_{\gamma,n+1}^2)(1+f_{\bar{\gamma},n}^2)}}c_{k,n+1,\gamma}^{\dagger}c_{k,n,\bar{\gamma}}\\\nonumber
&&+\sum_{k,n,\gamma,\bar{\gamma}}\frac{\gamma f_{\gamma,n}}{\sqrt{(1+f_{\gamma,n}^2)(1+f_{\bar{\gamma},n+1}^2)}}c_{k,n,\gamma}^{\dagger}c_{k,n+1,\bar{\gamma}}\Big\}\\\nonumber
&&J_y=-v_F\Big\{\sum_{k,\gamma}\frac{-i}{\sqrt{1+f_{\gamma,1}^2}}(c_{k,0}^{\dagger}c_{k,1,\gamma}-c_{k,1,\gamma}^{\dagger} c_{k,0})
+\sum_{k,n,\gamma,\bar{\gamma}}\frac{-i\bar{\gamma}f_{\bar{\gamma},n}}{\sqrt{(1+f_{\gamma,n+1}^2)(1+f_{\bar{\gamma},n}^2)}}c_{k,n+1,\gamma}^{\dagger}c_{k,n,\bar{\gamma}}\\\nonumber
&&+\sum_{k,n,\gamma,\bar{\gamma}}\frac{i\gamma f_{\gamma,n}}{\sqrt{(1+f_{\gamma,n}^2)(1+f_{\bar{\gamma},n+1}^2)}}c_{k,n,\gamma}^{\dagger}c_{k,n+1,\bar{\gamma}}\Big\}
\end{eqnarray}
 The particle current current correlation $L^{11}_{xx}(\omega)$ and $L^{11}_{xy}(\omega)$ from Eq.(\ref{kubo}) are given by
\begin{eqnarray}\nonumber   
&&L_{xx}^{11}(\omega,T,\mu)=\frac{-T v_F^2}{\pi l_B^2\hbar}\frac{1}{\omega}\int_{-\infty}^{\infty}\frac{d\epsilon}{2\pi}[n_F(\epsilon+\omega)-n_F(\epsilon)]\Big[\sum_{\gamma}\frac{1}{1+f_{\gamma,1}^2}[\Im G(0;\epsilon+0^+)\Im G(1,\gamma;\epsilon+\omega+0^+)\\\nonumber&&+\Im G(1,\gamma;\epsilon+0^+)\Im G(0;\epsilon+\omega+0^+)]+\sum_{n,\gamma,\bar{\gamma}}\frac{f_{\bar{\gamma},n}^2}{(1+f_{\gamma,n+1}^2)(1+f_{\bar{\gamma},n}^2)}\Im G(n+1,\gamma;\epsilon+0^+)\\\nonumber&&\times\Im G(n,\bar{\gamma};\epsilon+\omega+0^+)
+\sum_{n,\gamma,\bar{\gamma}}\frac{f_{\gamma,n}^2}{(1+f_{\bar{\gamma},n+1}^2)(1+f_{\gamma,n}^2)}\Im G(n,\gamma;\epsilon+0^+)\Im G(n+1,\bar{\gamma};\epsilon+\omega+0^+)\\\nonumber
&&+\sum_{\gamma}\frac{\Im G(0;\epsilon+0^+)\Im G(1,\gamma,-\gamma;\epsilon+\omega+0^+)+\Im G(1,\gamma,-\gamma;\epsilon+0^+)\Im G(0;\epsilon+\omega+0^+)}{\sqrt{(1+f_{\gamma,1}^2)(1+f_{-\gamma,1}^2)}}\\\nonumber
&&+\sum_{n,\gamma,\bar{\gamma}}\frac{f_{\bar{\gamma},n}^2}{(1+f_{\bar{\gamma},n}^2)\sqrt{(1+f_{\gamma,n+1}^2)(1+f_{-\gamma,n+1}^2)}}\Im G(n+1,\gamma,-\gamma;\epsilon+0^+)\Im G(n,\bar{\gamma};\epsilon+\omega+0^+)\\\nonumber
&&-\sum_{n,\gamma,\bar{\gamma}}\frac{f_{\bar{\gamma},n}f_{-\bar{\gamma},n}}{(1+f_{\gamma,n+1}^2)\sqrt{(1+f_{\bar{\gamma},n}^2)(1+f_{-\bar{\gamma},n}^2)}}\Im G(n+1,\gamma;\epsilon+0^+)\Im G(n,\bar{\gamma},-\bar{\gamma};\epsilon+\omega+0^+)\\\nonumber
&&+\sum_{n,\gamma,\bar{\gamma}}\frac{f_{\gamma,n}^2}{(1+f_{\gamma,n}^2)\sqrt{(1+f_{\bar{\gamma},n+1}^2)(1+f_{-\bar{\gamma},n+1}^2)}}\Im G(n,\gamma;\epsilon+0^+)\Im G(n+1,\bar{\gamma},-\bar{\gamma};\epsilon+\omega+0^+)\\\label{l11xx}
&&-\sum_{n,\gamma,\bar{\gamma}}\frac{f_{\gamma,n}f_{-\gamma,n}}{(1+f_{\bar{\gamma},n+1}^2)\sqrt{(1+f_{\gamma,n}^2)(1+f_{-\gamma,n}^2)}}\Im G(n,\gamma,-\gamma;\epsilon+0^+)\Im G(n+1,\bar{\gamma};\epsilon+\omega+0^+)\Big]
\end{eqnarray}
\begin{eqnarray}\nonumber
&&L_{xy}^{11}(\omega,T,\mu)=\frac{T v_F^2}{2\pi l_B^2\hbar}\frac{1}{\omega}\int_{-\infty}^{\infty}\frac{d\epsilon}{2\pi}\tanh(\frac{\epsilon}{2T})\Big[\sum_{\alpha=\pm1,\gamma}\frac{\alpha}{1+f_{\gamma,1}^2}[\Re G(1,\gamma;\epsilon+\alpha\omega+0^+)\Im G(0;\epsilon+0^+)\\\nonumber
&&-\Im G(1,\gamma;\epsilon+0^+)\Re G(0;\epsilon+\alpha\omega+0^+)]+\sum_{n,\alpha=\pm1,\gamma,\bar{\gamma}}\frac{\alpha f_{\bar{\gamma},n}^2}{(1+f_{\gamma,n+1}^2)(1+f_{\bar{\gamma},n}^2)}\Re G(n+1,\gamma;\epsilon+\alpha\omega+0^+)\\\nonumber
&&\times\Im G(n,\bar{\gamma};\epsilon+0^+)
-\sum_{n,\alpha=\pm1,\gamma,\bar{\gamma}}\frac{\alpha f_{\gamma,n}^2}{(1+f_{\bar{\gamma},n+1}^2)(1+f_{\gamma,n}^2)}\Re G(n,\gamma;\epsilon+\alpha\omega+0^+)\Im G(n+1,\bar{\gamma};\epsilon+0^+)\\\nonumber
&&+\sum_{\alpha=\pm 1,\gamma}\frac{\alpha[\Re G(1,\gamma,-\gamma;\epsilon+\alpha\omega+0^+)\Im G(0;\epsilon+0^+)-\Im G(1,\gamma,-\gamma;\epsilon+0^+)\Re G(0;\epsilon+\alpha\omega+0^+)]}{\sqrt{(1+f_{\gamma,1}^2)(1+f_{-\gamma,1}^2)}}\\\nonumber
&&+\sum_{n,\alpha=\pm 1,\gamma,\bar{\gamma}}\frac{\alpha f_{\bar{\gamma},n}^2}{(1+f_{\bar{\gamma},n}^2)\sqrt{(1+f_{\gamma,n+1}^2)(1+f_{-\gamma,n+1}^2)}}\Re G(n+1,\gamma,-\gamma;\epsilon+\alpha\omega+0^+)\Im G(n,\bar{\gamma};\epsilon+0^+)\\\nonumber
&&-\sum_{n,\alpha=\pm 1,\gamma,\bar{\gamma}}\frac{\alpha f_{\bar{\gamma},n}f_{-\bar{\gamma},n}}{(1+f_{\gamma,n+1}^2)\sqrt{(1+f_{\bar{\gamma},n}^2)(1+f_{-\bar{\gamma},n}^2)}}\Re G(n+1,\gamma;\epsilon+\alpha\omega+0^+)\Im G(n,\bar{\gamma},-\bar{\gamma};\epsilon+0^+)\\\nonumber
&&-\sum_{n,\alpha=\pm 1,\gamma,\bar{\gamma}}\frac{\alpha f_{\gamma,n}^2}{(1+f_{\gamma,n}^2)\sqrt{(1+f_{\bar{\gamma},n+1}^2)(1+f_{-\bar{\gamma},n+1}^2)}}\Re G(n,\gamma;\epsilon+\alpha\omega+0^+)\Im G(n+1,\bar{\gamma},-\bar{\gamma};\epsilon+0^+)\\\label{l11xy}
&&+\sum_{n,\alpha=\pm 1,\gamma,\bar{\gamma}}\frac{\alpha f_{\gamma,n}f_{-\gamma,n}}{(1+f_{\bar{\gamma},n+1}^2)\sqrt{(1+f_{\gamma,n}^2)(1+f_{-\gamma,n}^2)}}\Re G(n,\gamma,-\gamma;\epsilon+\alpha\omega+0^+)\Im G(n+1,\bar{\gamma};\epsilon+0^+)\Big]
\end{eqnarray}
The sum over $k$ (in $y$ direction) results in the normalization factor $N\bar{V}$ being replaced by $2\pi l_B^2$.
\end{widetext}


\begin{thebibliography}{999}
\bibliographystyle{unstr}
\bibitem{Kane}J. C. Y. Teo, L. Fu, and C. L. Kane, Phys. Rev. {\bf B 78}, 045426 (2008); L. Fu, C. L. Kane, and E. J. Mele, Phys. Rev. Lett. {\bf 98}, 106803 (2007).
\bibitem{Roy} R. Roy, Phys. Rev. {\bf B 79}, 195322 (2009); J. E. Moore and L. Balents, Phys. Rev. {\bf B 75}, 121306(R) (2007).
\bibitem{Fang} H.-J. Zhang, C.-X. Liu, X.-L. Qi, X.-Y. Deng, X. Dai, S.-C. Zhang, and Z. Fang, Phys. Rev. {\bf B 80}, 085307 (2009).
\bibitem{Hsieh} D. Hsieh, D. Qian, L. Wray, Y. Xia, Y. Hor, R. J. Cava, and. M. Z. Hasan, Nature (London) {\bf 452}, 970 (2008).
\bibitem{Xia}Y. Xia, L. Wray, D. Qian, D. Hsieh, A. Pal, H. Lin, A. Bansil, D. Grauer, Y.S. Hor, R.J. Cava, M.Z. Hasan, ArXiv:0812.2078.
\bibitem{Zhang} M. K$\ddot o$nig, S. Wiedmann, C Br$\ddot u$ne, A. Roth, H. Buhmann, L. W. Molenkamp, X.-L. Qi, and S.-C. Zhang, Science {\bf 381}, 766 (2007); Y. Xia, D. Qian, D. Hsieh, L. Wray, A. Pal, H. Lin, A. Bansil, D. Grauer, Y. S. Hor, R. J. Cava, and M. Z. Hasan, Nature Phys. {\bf 5}, 398 (2009). 
\bibitem{steinberg1}H. Steinberg, D.R. Gardner, Y.S. Lee, and P. Jarillo-Herrero, Nano Lett. {\bf 10}, 5032 (2010).
\bibitem{analytis1}J.G. Analytis, J-H Chu, Y. Chen, F. Corredor, R.D. McDonald, Z.X. Shen, and I.R. Fisher, Phys. Rev. B  {\bf 81}, 205407 (2010)
\bibitem{check1}J.G. Checkelsky, Y.S. Hor, M.-H. Liu, D.-X. Qu, R.J. Cava, and N.P. Ong, Phys. Rev. Lett. {\bf 103}, 246601 (2009).
\bibitem{wei1}P. Wei, Z. Wang, X. Liu, V. Aji, and J. Shi, unpublished
\bibitem{jchen}J. Chen, H.J.  Qin, F. Yang, J. Liu, T. Guan, F.M.  Qu, G.H. Zhang, J.R. Shi, X.C. Xie, C.L. Yang, K.H.  Wu, Y.Q.  Li, and L. Lu, Phys. Rev. Lett. {\bf 105}, 176602 (2010).
\bibitem{Ong}D.-X. Qu, Y. S. Hor, J. Xiong, R. J. Cava, and N. P. Ong, Science {\bf 329}, 821 (2010).
\bibitem{Mondal} S. Mondal, D. Sen, K. Sengupta, and R. Shankar, Phys. Rev. {\bf B 82}, 045120 (2010).
\bibitem{Jonson} M. Jonson and S. M. Girvin, Phys. Rev. {\bf B 29}, 1939 (1984); S. M. Girvin and M. Jonson, J. Phys. C {\bf 15}, L1147 (1982).
\bibitem{Zuev}Y. M. Zuev, W. Chang, and P. Kim, Phys. Rev. Lett. {\bf 102}, 096807 (2009).
\bibitem{Check}J. G. Checkelsky and N. P. Ong, Phys. Rev. B {\bf 80}, 081413(R) (2009).
\bibitem{Peres} N. M. R. Peres, F. Guinea, and A. H. Castro Neto, Phys. Rev. {\bf B 73}, 125411 (2006).
\bibitem{Ando} Y. Zheng and T. Ando, Phys. Rev. {\bf B 65}, 245420 (2002).
\bibitem{Ugarte}V. Ugarte, V. Aji and C.M. Varma, ArXiv:1007.3533, unpublished.
\bibitem{Hwang} E. H. Hwang, E. Rossi, S. Das Sarma, Phys. Rev. {\bf B 80}, 235415 (2009).
\bibitem{Dassarma}S. Das Sarma, S. Adam, E.H. Hwang, and E. Rossi, Rev. Mod. Phys. {\bf 83}, 407 (2011).
\bibitem{Yan} X. Z. Yan, Y. Romiah, and C. S. Ting, Phys. Rev. {\bf B 80}, 165423 (2009). 
\bibitem{Wei}P. Wei, W. Bao, Y. Pu, C. N. Lau, and J. Shi, Phys. Rev. Lett. {\bf 102}, 166808 (2009).
\bibitem{Bruus} H. Bruus and K. Flesberg, \textsl{Many-body Quantum Theory in Condensed Matter Physics} (Oxford University Press, Oxford, 2004).
\bibitem{Zhu} L. Zhu, R. Ma, L. Sheng, M. Liu, and D-N Sheng, Phys. Rev. Lett. {\bf 104}, 076804 (2010).
\bibitem{Hao}L. Hao and  T. K. Lee, Phys. Rev. {\bf B 82}, 245415(2010).
\bibitem{Zaliznyak} L. Zhang, Y. Zhang, M. Khodas, T. Valla, and I. A. Zaliznyak, Phys. Rev. Lett. {\bf 105}, 046804 (2010).
\bibitem{Adam}S. Adam, E. H. Hwang, V. M. Galitski, and S. D. Sarma, Proc. Natl. Acad. Sci. U.S.A. {\bf 104}, 18392 (2007).
\bibitem{cho}S. Cho and M. S. Fuhrer, Phys. Rev. {\bf B 77}, 081402(R) (2008).
\bibitem{Pchen}P. Cheng, C. Song, T. Zhang, Y. Zhang,Y.  Wang, J.-F. Jia, J. Wang, Y. Wang, B.-F Zhu, X. Chen, X. Ma, K. He, L. Wang, X. Dai, Z. Fang, X. Xie, X.-L. Qi, C.-X. Liu, S.C.  Zhang, and Q.-K. Xue, Phys. Rev. Lett. {\bf 105}, 076801 (2010).
\bibitem{Hanaguri}T. Hanaguri, K.  Igarashi, M. Kawamura, H. Takagi, and T. Sasagawa, Phys. Rev. B {\bf 82}, 081305 (2010).
\end{thebibliography}
\end{document}